\documentclass[amsmath,showpacs,nofootinbib,12pt]{revtex4-2}
\usepackage{graphicx}
\usepackage{dcolumn}
\usepackage{bm}
\usepackage{color} 
\usepackage{slashed}
\begin{document}
\newcommand{\hs}{\hspace*{0.5cm}}
\newcommand{\vs}{\vspace*{0.5cm}}
\newcommand{\be}{\begin{equation}}
\newcommand{\ee}{\end{equation}}
\newcommand{\bea}{\begin{eqnarray}}
\newcommand{\eea}{\end{eqnarray}}
\newcommand{\ben}{\begin{enumerate}}
\newcommand{\een}{\end{enumerate}}
\newcommand{\bde}{\begin{widetext}}
\newcommand{\ede}{\end{widetext}}
\newcommand{\nn}{\nonumber}
\newcommand{\crn}{\nonumber \\}
\newcommand{\Tr}{\mathrm{Tr}}
\newcommand{\non}{\nonumber}
\newcommand{\noi}{\noindent}
\newcommand{\al}{\alpha}
\newcommand{\la}{\lambda}
\newcommand{\bet}{\beta}
\newcommand{\ga}{\gamma}
\newcommand{\va}{\varphi}
\newcommand{\om}{\omega}
\newcommand{\pa}{\partial}
\newcommand{\+}{\dagger}
\newcommand{\fr}{\frac}
\newcommand{\bc}{\begin{center}}
\newcommand{\ec}{\end{center}}
\newcommand{\Ga}{\Gamma}
\newcommand{\de}{\delta}
\newcommand{\De}{\Delta}
\newcommand{\ep}{\epsilon}
\newcommand{\varep}{\varepsilon}
\newcommand{\ka}{\kappa}
\newcommand{\La}{\Lambda}
\newcommand{\si}{\sigma}
\newcommand{\Si}{\Sigma}
\newcommand{\ta}{\tau}
\newcommand{\up}{\upsilon}
\newcommand{\Up}{\Upsilon}
\newcommand{\ze}{\zeta}
\newcommand{\ps}{\psi}
\newcommand{\Ps}{\Psi}
\newcommand{\ph}{\phi}
\newcommand{\vph}{\varphi}
\newcommand{\Ph}{\Phi}
\newcommand{\Om}{\Omega}
\newcommand{\AdrHEPC}{Phenikaa Institute for Advanced Study and Faculty of Basic Science, Phenikaa University, Yen Nghia, Ha Dong, Hanoi 100000, Vietnam}

\title{Physics implication from higher weak isospin decomposition} 

\author{Phung Van Dong} 
\email{dong.phungvan@phenikaa-uni.edu.vn (corresponding author)}
\author{Duong Van Loi}
\email{loi.duongvan@phenikaa-uni.edu.vn}
\affiliation{\AdrHEPC} 
\date{\today}

\begin{abstract}

The $SU(3)_L\otimes U(1)_X$ symmetry actually studied is directly broken to the electroweak symmetry $SU(2)_L\otimes U(1)_Y$ by a Higgs triplet, predicting a relevant new physics at TeV scale. This work argues, by contrast, that the higher weak isospin $SU(3)_L$ might be broken at a high energy scale, much beyond $1$ TeV, by a Higgs octet to an intermediate symmetry $SU(2)_L\otimes U(1)_{T_8}$ at TeV, before the latter $U(1)_{T_8}$ recombined with $U(1)_X$ defines (i.e., broken to) $U(1)_Y$ by a Higgs singlet. The new physics coupled to $SU(3)_L$ breaking phase is decoupled, whereas what remains is a novel family-nonuniversal abelian model, $U(1)_{T_8}\otimes U(1)_X$, significantly overhauling the standard model as well as yielding consistent results for neutrino mass, dark matter, $W$-mass anomaly, and FCNC, differently from the usual 3-3-1 model.       

\end{abstract}

\maketitle

\section{Introduction}

The experimental evidences of neutrino oscillation \cite{Kajita:2016cak,McDonald:2016ixn} and dark matter existence \cite{Jungman:1995df,Bertone:2004pz,Arcadi:2017kky} indicate that the standard model must be extended. Recently, the $W$-boson mass anomaly measured by CDF collaboration with high precision at 7$\sigma$ also requires an extension for the standard model \cite{CDF:2022hxs}. On the theoretical side, the standard model cannot explain the quantization of electric charge and the existence of just three families of fermions.   

Among compelling attempts that address the above questions, the $SU(3)_C\otimes SU(3)_L\otimes U(1)_X$ (called 3-3-1) model is well-motivated due to its ability to predict the number of families to match that of colors by anomaly cancellation condition \cite{331v1,331v2,331pp,331f,331flt}. Further, the electric charge quantization in the 3-3-1 model naturally arises from specific fermion contents under gauge symmetry principles \cite{ecq1,ecq2,ecq3,ecq4,ecq5}. The 3-3-1 model can provide both appropriate neutrino masses by implementing seesaw mechanisms \cite{neu1,neu2,neu3,neu4,neu5,neu6,neu7,neu8,neu9,neu10,neu11,neu12,neu13} and dark matter stability by supplementing global or residual gauge symmetry \cite{d1,d2,d3,d4,d5,d6,d7,d8,d9,d10,d11,d12,d13,d14,d15,d16,d17,d18}. Recently, the 3-3-1 model has been shown to give a suitable answer for the $W$-mass deviation \cite{wm331}.  

The strong interest of the 3-3-1 model is perhaps associated with spontaneous symmetry breaking sector, which leads to distinct kinds of new physics. The minimal, but nontrivial, representations of Higgs fields under 3-3-1 symmetry must contain a triplet, a sextet, and an octet, as well as their complex conjugates if any. The triplet and sextet are eventually studied, since they directly couple to fermions, providing relevant fermion masses, besides responsible for breaking the gauge symmetry. Hence, in the literature, the symmetry breaking scheme that breaks 3-3-1 symmetry directly down to the standard model due to such Higgs fields is uniquely recognized. By contrast, the octet---which possesses a representation as of the adjoint gauge field---does not couple to fermions in Yukawa Lagrangian.\footnote{This is obviously opposite to the similar situation of a weak-isospin scalar triplet in the standard model, which can couple to leptons.} Hence, such octet has eventually been suppressed. 

This work considers all of the Higgs effects including the octet, indicating that there might be two new physics phases for the 3-3-1 model, which has not been observed/recognized before. Indeed, the octet breaks the 3-3-1 symmetry down to an intermediate symmetry that contains family-dependent abelian charges, whereas the triplet/sextet has a singlet component, breaking the intermediate symmetry to the standard model symmetry. Interestingly, given that the high symmetry phase broken by the octet is decoupled, the intermediate abelian theory supplies interesting physical results for neutrino mass and dark matter, as well as recent physical issues, which are not presented in the usual 3-3-1 model. 

The rest of this work is organized as follows. In Sec. \ref{331}, we reconsider the 3-3-1 model when including the effect of a Higgs octet as well as imposing a matter parity that governs this setup. In Sec. \ref{fdphase}, we investigate the family-dependent abelian theory that results after the 3-3-1 new physics is decoupled, where the $W$-mass deviation and FCNC are discussed. The fermion mass including that of neutrinos and dark matter stability are investigated in the subsequent sections, Secs. \ref{fmass} and \ref{dms}, respectively. We summarize the results and make conclusion as well as outlook in the last section, Sec. \ref{con}.     

\section{\label{331} Revisiting the 3-3-1 model}

We first provide the necessary features of the 3-3-1 model, gauge symmetry and fermion content. We then revisit the 3-3-1 symmetry breaking with the presence of a Higgs octet. Lastly, we discuss $B-L$ symmetry and its resultant matter parity.   

\subsection{Gauge symmetry and fermion content}

In the standard model, the weak isospin $SU(2)_L$ forms left-handed fermions as isodoublets, $l_L=(\nu_L,e_L)$ and $q_L=(u_L,d_L)$, while putting right-handed partners, $e_R$, $u_R$, and $d_R$, as isosinglets. Additionally, the $[SU(2)_L]^3$ anomaly always vanishes, since $\mathrm{Tr}[\{T_i,T_j\}T_k]=0$ for every $SU(2)_L$ charge, $T_i$ $(i=1,2,3)$. Enlarging $SU(2)_L$ to a higher weak isospin, say $SU(3)_L$, the relevant $[SU(3)_L]^3$ anomaly does not vanish, i.e. $\mathrm{Tr}[\{T_m,T_n\}T_p]\neq 0$, for complex representation, where $T_m$ $(m=1,2,\cdots,8)$ denotes $SU(3)_L$ charge. This subsequently constrains fermion content by anomaly cancelation \cite{ano1,ano2,ano3,ano4}. Notice that the new symmetry must span $SU(3)_C\otimes SU(3)_L$ by imposing the color charge. 

The $SU(3)_L$ group has two fundamental representations, $3=2\oplus 1$ and $3^*=2^*\oplus 1$, as decomposed under $SU(2)_L$. Hence we enlarge known fermion isodoublets $(f_1,f_2)$ to respective triplets/antitriplets, i.e. $(f_1,f_2,f_s)\sim 3$ or $(f_2,-f_1,f_s)\sim 3^*$, where $(f_2,-f_1)\sim 2^*$ given that $(f_1,f_2)\sim 2$, isosinglets $f_s\sim 1$ necessarily added, while other known fermion isosinglets presumably transform as $SU(3)_L$ singlets by themselves. Obviously, the $[SU(3)_L]^3$ anomaly is nonzero for each family and cancelled over families if the number of triplets equals that of antitriplets, since the anomaly contributions of $3$ and $3^*$ are opposite, and note that the color number must be appropriately counted. It follows that family number is a multiple of color number, 3. Further, QCD asymptotic freedom limits family number less than or equal to 5, since each exotic quark isosinglet is added to a family that also contributes to the strong running coupling. Therefore, the family number matches 3 as observed. That said, the fermion content is arranged under $SU(3)_L$ as \be \psi_{aL}=\begin{pmatrix} \nu_{aL}\\ e_{aL}\\ N_{aL}\end{pmatrix}\sim 3\ee for three lepton families $a=1,2,3$, \be Q_{\al L}=\begin{pmatrix} d_{\al L}\\ -u_{\al L}\\ J_{\al L}\end{pmatrix} \sim 3^*\ee  for first two quark families $\al=1,2$, and \be Q_{3L}=\begin{pmatrix} u_{3L}\\ d_{3L}\\ J_{3L}\end{pmatrix}\sim 3\ee for third quark family, whereas relevant right-handed fermions $e_{aR}$, $N_{aR}$, $u_{aR}$, $d_{aR}$, and $J_{aR}$ are all put in $SU(3)_L$ singlets. Notice that minimal 3-3-1 versions assume either $N_L=(e_R)^c$ or $N_L=(\nu_R)^c$, while $N_R$ is suppressed. 

Theoretically, $N$ is an arbitrary new lepton. Let the electric charge of it be $q$, i.e. $Q(N)=q$. We have $Q=\mathrm{diag}(0,-1,q)$ for a lepton triplet, implying an algebra 
\bea \left[Q, T_1\pm i T_2\right] &=&\pm (T_1\pm i T_2),\\ 
\left[Q, T_4\pm i T_5\right] &=& \mp q(T_4\pm i T_5),\\  
\left[Q, T_6\pm i T_7\right] &=& \mp (1+q)(T_6\pm i T_7).\eea 
As the standard model, $Q$ does not commute with $SU(3)_L$ charge. With the aid of commutation relations for $T_{3,8}$ with $T_4\pm i T_5$ and $T_6\pm i T_7$, that is \bea && \left[T_3,T_1\pm i T_2\right]=\pm (T_1\pm i T_2),\ \left[T_8,T_1\pm i T_2\right]=0,\\ 
&& \left[T_3,T_4\pm i T_5\right]=\pm 1/2 (T_4\pm i T_5),\ \left[T_8,T_4\pm i T_5\right]=\pm \sqrt{3}/2(T_4\pm i T_5),\\
&& \left[T_3,T_6\pm i T_7\right]=\mp 1/2(T_6\pm i T_7),\ \left[T_8,T_6\pm i T_7\right]=\pm \sqrt{3}/2(T_6\pm i T_7), \eea one finds a new charge, \be X=Q-T_3-\beta T_8,\ee that commutes with every $T_i$ charge, i.e. $[X,T_i]=0$, given that  \be \beta=-\fr{1+2q}{\sqrt{3}}.\ee On the other hand, $Q$ does not algebraically close with $SU(3)_L$ charge, since otherwise algebraic closure requires $Q$ to be a linear composition of $T_i$, i.e. $Q=x_i T_i$, leading to $\mathrm{Tr}(Q)=0$, which is incorrect for present fermion representations. It follows that $X$ is an independent abelian charge, defining a symmetry,   
\be SU(3)_C\otimes SU(3)_L \otimes U(1)_X, \ee where the color group is also imposed, called 3-3-1. Comparing to the standard model, i.e. $Y=Q-T_3$, the hypercharge is identical to \be Y=\beta T_8+X.\ee It is easy to find electric charge of new quarks, $Q(J_3)= 2/3+q$ and $Q(J_{1,2})=-1/3-q$. The fermion content transforms under the 3-3-1 symmetry as 
\bea && \psi_{aL}\sim \left(1,3,\fr{q-1}{3}\right),\ e_{aR}\sim (1,1,-1),\ N_{aR}\sim (1,1,q),\\
&& Q_{\al L}\sim \left(3,3^*,-\fr{q}{3}\right),\ Q_{3L}\sim \left(3,3,\fr{q+1}{3}\right),\\
&& u_{aR}\sim (3,1,2/3),\ d_{aR}\sim (3,1,-1/3),\\
&& J_{\al R} \sim (3,1,-1/3-q),\ J_{3R}\sim (3,1,2/3+q).\eea

In the literature, there exists an alternative fermion content that is anomaly free, achieved by switching the current representations with conjugated representations, such as $(\nu_{aL},e_{aL},N^q_{aL})\rightarrow (e_{aL},-\nu_{aL},N^q_{aL})$, $(d_{\al L}, -u_{\al L}, J^{-1/3-q}_{\al L})\rightarrow (u_{\al L},d_{\al L}, J^{-1/3-q}_{\al L} )$, and $(u_{3L}, d_{3L},J^{2/3+q}_{3L})\rightarrow (d_{3L},-u_{3L},J^{2/3+q}_{3L})$, whereas the right-handed fermion partners retain the same. This correspondingly changes $\beta\rightarrow -\beta$ and leads to a model with rather similar phenomena, which is not considered in this work. Surely, our results apply for both types of fermion content. In what follows we consider 3-3-1 breaking.      

\subsection{Higher weak isospin decomposition and hypercharge realization}

In the literature, the 3-3-1 symmetry is broken by a Higgs triplet, \be \chi=\begin{pmatrix} \chi^{-q}_1\\ \chi^{-1-q}_2\\ \chi^0_3\end{pmatrix}\sim \left(1,3,\fr{-1-2q}{3}\right),\ee with vacuum expectation value (VEV), \be \langle \chi\rangle =\begin{pmatrix} 0\\ 0 \\ w/\sqrt{2}\end{pmatrix},\ee {\it directly} down to the standard model, that is \bc \begin{tabular}{c} 
$SU(3)_C\otimes SU(3)_L \otimes U(1)_X$ \\ $\downarrow w$ \\ $SU(3)_C\otimes SU(2)_L \otimes U(1)_Y$ \end{tabular}\ec This is because the new charges $T_{4,5,6,7,8}$ do not annihilate the vacuum, i.e. $T_{4,5,6,7,8}\langle \chi\rangle \neq 0$, and are thus broken, while the weak isospin conserves the vacuum, i.e. $T_{1,2,3}\langle \chi\rangle =0$. Additionally, $X\langle \chi\rangle =-\fr{1+2q}{3}\langle \chi\rangle \neq 0$ for $q\neq -1/2$ yields that $X$ is broken, but $Y$ conserves the vacuum similar to the weak isospin, since $Y\langle \chi \rangle =0$.  
Note that for $q=-1/2$, i.e. $\beta=0$, one has $X=Y$, hence the above statement is valid for every $q$, i.e. every 3-3-1 model. It is easily verified that a Higgs sextet designed for 3-3-1 breaking takes the same situation. As a matter of fact, the symmetry breaking by a Higgs triplet or sextet has been extensively studied, leading to a new physics consistently bounded at TeV scale.\footnote{See, for a recent constraint on the 3-3-1 model, \cite{queiroz}.} 

While such a new physics has not been confirmed, we would like to argue in this work that the 3-3-1 symmetry may be broken at a very high energy scale, much beyond TeV scale, but imprinted at TeV is an abelian theory entirely different from the usual 3-3-1 model. That said, the 3-3-1 symmetry is broken at high energy by a Higgs octet, 
\be S=\sqrt{2} T_m S_m = \begin{pmatrix}\fr{S_3}{\sqrt{2}}+\fr{S_8}{\sqrt{6}} & S^+_{12} & S^{-q}_{13}\\
S^-_{12} & -\fr{S_3}{\sqrt{2}}+\fr{S_8}{\sqrt{6}} & S^{-1-q}_{23}\\
S^q_{13} & S^{1+q}_{23} & -\fr{2S_8}{\sqrt{6}}
\end{pmatrix}\sim (1,8,0),\label{sstt23l}\ee which transforms the same with the $SU(3)_L$ gauge field in adjoint representation. Notice that in this case the $SU(3)_L$ generators are related to the Gell-Mann matrices, such as $T_m=\la_m/2$ ($m=1,2,3,\cdots,8$), while the non-Hermitian scalars  
\be S^\pm_{12}=\fr{S_1\mp i S_2}{\sqrt{2}},\ S^{\mp q}_{13}=\fr{S_4 \mp i S_5}{\sqrt{2}},\ S^{\mp(1+q)}_{23}=\fr{S_6\mp i S_7}{\sqrt{2}}\ee are associated with the weight-raising and lowering operators, $(T_1\pm i T_2)/\sqrt{2}$, $(T_4\pm i T_5)/\sqrt{2}$, and $(T_6\pm i T_7)/\sqrt{2}$, respectively. Notice also that each complex field $S_{12}$, $S_{13}$, and $S_{23}$ contains two independent degrees of freedom as its real and imaginary parts originating from $S_m$; or in other words, as appeared in (\ref{sstt23l}), $S^+_{12}$ and $S^-_{12}\equiv (S^+_{12})^*$ and so forth for pairs of $S_{13}$ and $S_{23}$ are independent degrees of freedom. When $S_8$ develops a VEV, i.e. $\langle S_8 \rangle =\La$, the Higgs octet has a VEV structure,
\be \langle S \rangle = \fr{1}{\sqrt{6}}\begin{pmatrix}
\La & 0 & 0 \\
0 & \La & 0\\
0 & 0 & -2 \La\end{pmatrix}. 
\ee  Such a $S$ breaks the 3-3-1 symmetry to an intermediate symmetry, 
 \bc \begin{tabular}{c} 
$SU(3)_C\otimes SU(3)_L \otimes U(1)_X$ \\ $\downarrow \La $ \\ $SU(3)_C\otimes SU(2)_L \otimes U(1)_{T_8}\otimes U(1)_X$ \end{tabular}\ec Indeed, only $T_{4,5,6,7}$ are broken due to $[T_{4,5,6,7},\langle S\rangle]\neq 0$, whereas $T_{1,2,3}$, $T_8$, and $X$ are all conserved by $\La$, because of \be [T_{1,2,3,8},\langle S \rangle ]=0=[X,\langle S \rangle],\ \ee where note that $T_m=\la_m/2$, as mentioned. That said, the octet breaks $SU(3)_L\rightarrow SU(2)_L\otimes U(1)_{T_8}$, while conserving $U(1)_X$, which differs from the cases of Higgs triplet and sextet. Interestingly, this kind of breaking leads to a decomposition of every particle multiplet of $SU(3)_L$ into those of $SU(2)_L\otimes U(1)_{T_8}$ below the breaking energy scale, for instance, the isodoublet-isosinglet splitting in $3=2\oplus 1$ or in $3^*=2^*\oplus 1$ in which $T_8$ values of component $2$ or $2^*$ and $1$ are appropriately taken. Generically, a $SU(3)_L$ multiplet separates into a normal isomultiplet---a weak isospin multiplet that has normal $B-L$ number as of the standard model---simply called {\it isomultiplet} and an abnormal isomultiplet---a weak isospin multiplet that has abnormal $B-L$ number as of dark sector---called {\it isopartner} (see below for imposing $B-L$ and matter parity to this model).   

Similar to the Higgs octet, the $SU(3)_L$ gauge field $A=\sqrt{2}T_m A_m$ can be written in terms of non-Hermitian gauge fields, $W^\pm=(A_1\mp i A_2)/\sqrt{2}$, $U^{\mp q}=(A_4\mp i A_5)/\sqrt{2}$, and $V^{\mp (1+q)}=(A_6\mp i A_7)/\sqrt{2}$, coupled to the weight-raising and lowering operators, $(T_1\mp i T_2)/\sqrt{2}$, $(T_4\pm i T_5)/\sqrt{2}$, and $(T_6\pm i T_7)/\sqrt{2}$, besides neutral gauge fields, $A_3$ and $A_8$, coupled to $T_3$ and $T_8$, respectively. Given that the 3-3-1 breaking scale is very high, say $\La\gg 1$ TeV, the new gauge bosons $U^{\mp q}$ and $V^{\mp(1+q)}$ associated with broken charges $T_{4,5,6,7}$ obtain a large mass proportional to $\La$, which can be integrated out. Correspondingly, this gauges away the massless Goldstone bosons $S^{\mp q}_{13}$ and $S^{\mp(1+q)}_{23}$ as eaten by $U^{\mp q}$ and $V^{\mp(1+q)}$, respectively. The remaining components of $S$ all obtain a heavy mass at $\La$ scale, which must be integrated out too. What imprints at TeV is a novel abelian theory, $U(1)_{T_8}\otimes U(1)_X$, besides the usual group $SU(3)_C\otimes SU(2)_L$, as in the above breaking scheme due to the Higgs octet. Hence, in this 3-3-1 breaking phase, due to $SU(2)_L \otimes U(1)_{T_8}$ decomposition the new fields $N$, $J$, $\chi_3$, $A_8$, and $U(1)_X$ gauge field, called $B$, coexist but essentially separated from the standard model fields. Also, although unified at $\La$ scale, the $SU(2)_L$ and $U(1)_{T_8}$ couplings at TeV would be different due to the running coupling. As we will be seen, this new abelian theory is very predictive than the usual 3-3-1 model at TeV. Last, but not least, the isosinglet $\chi_3$ has both $T_8$ and $X$ charges. Hence, it breaks these charges down to the hypercharge, i.e. 
\bc \begin{tabular}{c} 
$SU(3)_C\otimes SU(2)_L \otimes U(1)_{T_8}\otimes U(1)_X$ \\ $\downarrow w $ \\ $SU(3)_C\otimes SU(2)_L \otimes U(1)_Y$ \end{tabular}\ec where $Y=\beta T_8+X$ is as given, while $w=\sqrt{2}\langle \chi_3\rangle$ is at TeV scale.  

In summary, there may be two new physics phases for the 3-3-1 model, such as 
\bc \begin{tabular}{c} 
$SU(3)_C\otimes SU(3)_L \otimes U(1)_X$ \\ $\downarrow \La $ \\ $SU(3)_C\otimes SU(2)_L \otimes U(1)_{T_8}\otimes U(1)_X$\\ $\downarrow w $ \\ $SU(3)_C\otimes SU(2)_L \otimes U(1)_Y$  \end{tabular}\ec The first phase is governed by a Higgs octet $S$, while the second phase close to the standard model is set by a Higgs singlet $\chi_3$. Given that $\La$ is large enough, i.e. $\La\gg w \sim $ TeV, only the second phase with abelian symmetry $U(1)_{T_8}\otimes U(1)_X$ affects the standard model, which must be probed in detail. Notice also that in the second phase every isomultiplet has its isopartner resulting from the higher weak isospin splitting according to $SU(3)_L\rightarrow SU(2)_L\otimes U(1)_{T_8}$, except for original $SU(3)_L$ singlets. For instance, the isopartner of $\chi^0_3$ is the isodoublet $(\chi^{-q}_1,\chi^{-1-q}_2)$, a potential candidate for dark matter since it is not eaten by $(U^{-q},V^{-1-q})$ gauge isodoublet unlike the usual 3-3-1 model. Additionally, normal Higgs isodoublets $(\eta^0_1,\eta^-_2)$ and $(\rho^+_1,\rho^0_2)$ necessary for breaking $SU(2)_L\otimes U(1)_Y\rightarrow U(1)_Q$ and generating appropriate masses have isopartners as $\eta^q_3$ and $\rho^{1+q}_3$ isosinglets, respectively, other potential candidates for dark matter. The new lepton isosinglet $N^q$ is the isopartner of $(\nu,e)$ isodoublet, possibly contributing to dark matter too, whereas the exotic quark isosinglet $J$---the isopartner of $(u,d)$ isodoublet---does not. All isomultiplet and corresponding isopartner are collected in Table \ref{tab0} for convenience in reading.  

\begin{table}[h]
\bc
\begin{tabular}{lccccccc}
\hline\hline
$SU(3)_L$ multiplet & Lepton & Quark & Gauge & & Higgs  & & Heavy-Higgs \\ 
\hline 
Isomultiplet: $P=1$  & $\begin{pmatrix}\nu\\
e \end{pmatrix}$ & $\begin{pmatrix} u\\
d \end{pmatrix}$ & $\begin{pmatrix} A_1\\ A_2 \\ A_3\end{pmatrix}\oplus A_8$ & $\begin{pmatrix}\eta_1 \\
\eta_2 \end{pmatrix}$ & $\begin{pmatrix}\rho_1 \\
\rho_2 \end{pmatrix}$ & $\chi_3$ & $\begin{pmatrix} S_1\\ S_2 \\ S_3\end{pmatrix}\oplus S_8$\\
Isopartner: $P=P^{\pm}$ & $N$ & $J$ & $\begin{pmatrix} U \\
V \end{pmatrix}$ & $\eta_3$ & $\rho_3$ & $\begin{pmatrix}\chi_1\\
\chi_2 \end{pmatrix}$  & $\begin{pmatrix} S_{13} \\
S_{23} \end{pmatrix}$ \\
\hline\hline
\end{tabular}
\caption[]{\label{tab0} Isopartner of normal/known isomultiplet}
\ec
\end{table}   

\subsection{$B-L$ symmetry and matter parity}

Before proceeding with the $U(1)_{T_8}\otimes U(1)_X$ model, let us examine the symmetry of baryon number minus lepton number $B-L$ and its residual matter parity $P$, which governs the current setup. By the same argument for $Q$, we can determine the behavior of $B-L$. Indeed, since the new leptons $N_a$ are arbitrary, let their $B-L$ be $n$, i.e. $[B-L](N_a)=n$. One has $B-L=\mathrm{diag}(-1,-1,n)$ for lepton triplets $(\nu_{aL},e_{aL},N_{aL})$, which generically neither commutes nor closes with $SU(3)_L$ charges, because of \bea \left[B-L, T_4\pm i T_5\right] &=& \mp (1+n)(T_4\pm i T_5),\\  
\left[B-L, T_6\pm i T_7\right] &=& \mp (1+n)(T_6\pm i T_7),\eea and $\mathrm{Tr}(B-L)\neq 0$ for present fermion representations. This $B-L$ assignment is motivated by the traditional 3-3-1 models with $N_L=(\nu_R)^c$ or $(e_R)^c$ for which $B-L=\mathrm{diag}(-1,-1,1)$ for lepton triplets and is also inspired by a TeV-scale seesaw mechanism with $L(N_L)=0$ \cite{mazeronu} for which $B-L=\mathrm{diag}(-1,-1,0)$ for relevant lepton triplets. Exceptionally, only if $N$ is a new lepton with $n=-1$, $B-L$ commutes with $SU(3)_L$ and the following trick does not apply, similar to the standard model. It is stressed that the 3-3-1 model also conserves $B+L$ but experiences anomalies when $B+L$ is promoted as a gauge symmetry, in contrast to $B-L$, which is not of interest. Using the given commutation relations, we derive an independent abelian charge \be M=B-L-\beta' T_8,\label{blg}\ee given that $\beta'=-2(1+n)/\sqrt{3}$. Since $M$ and $X$ are linearly independent as $B-L$ and $Q$ are, this defines a symmetry $SU(3)_C\otimes SU(3)_L\otimes U(1)_X\otimes U(1)_M$, called 3-3-1-1 \cite{d13,mp2,d17}. Acting $B-L=\beta' T_8+M$ on relevant multiplets, we derive $B-L$ for exotic particles, apart from that for new leptons as given, namely $[B-L](J_\al)=-2/3-n$, $[B-L](J_3)=4/3+n$, $[B-L](\chi_{1,2})=[B-L](\eta^*_3,\rho^*_3)=[B-L](S_{13},S_{23})=[B-L](U,V)=-1-n$, whereas other new bosons have $B-L=0$ like those of the standard model. Simultaneously, we obtain $M$-charge for $SU(3)_L$ multiplets, $M(\psi_{aL})=(n-2)/3$, $M(Q_{\al L})=-n/3$, $M(Q_{3L})=(n+2)/3$, $M(\chi)=-2(1+n)/3$, $M(\eta,\rho)=(1+n)/3$, and $M(S,A,B)=0$, while for right-handed fermion singlets $M$ coincides with $B-L$. 

Since $T_8$ is gauged, $M$ (thus $B-L$) must be gauged due to the relation (\ref{blg}), a result of noncommutative $B-L$. As proved in \cite{d13,mp2,d17}, the $U(1)_M$ physics (i.e., its gauge boson, Higgs field, and $\nu_{aR}$) may be decoupled at GUT scale, or at least above the Higgs octet VEV, leaving the 3-3-1 symmetry besides a residual matter parity conserved, such as \be P=(-1)^{3(B-L)+2s}=(-1)^{3(\beta' T_8+M)+2s},\ee where $s$ is spin. Note that the $S$, $\chi$, $\eta$, and $\rho$ vacua always conserve $B-L$ and $P$. Isomultiplets that possess a normal $B-L$ number (i.e., independent of $n$), including the standard model fields, $\eta_{1,2}$, $\rho_{1,2}$, $\chi_3$, $A_8$, $B$, and $S_{1,2,3,8}$, all have $P=1$, whereas their isopartners that have an abnormal $B-L$ number (i.e., dependent on $n$), including $N_a$, $J_a$, $\chi_{1,2}$, $\eta_3$, $\rho_3$, $S_{13,23}$, and $U,V$, all possess a nontrivial $P=P^{\pm}\equiv (-1)^{\pm(3n+1)}\neq 1$, which is odd, i.e. $P=-1$, for $n=2k/3=0,\pm 2/3,\pm 4/3,\pm 2,\cdots$, for which the matter parity of isomultiplet and isopartner are indicated in Table \ref{tab0}. Isopartner of normal isomultiplet is $P$-nontrivial, since their $B-L$'s differ by an amount depending on $n$ ($\beta'$), caused by $T_8$ according to (\ref{blg}). 

It is stressed that since $P$ is always conserved, the lightest isopartner---the lightest abnormal $B-L$ particle---is stabilized, responsible for dark matter (cf. \cite{mp2}). As specified, dark matter candidates include $N$, $\eta_3$, $\rho_3$, and $\chi_{1,2}$ depending on $q$ since they must be electrically neutral too. Note that since the dark matter candidate must be colorless and stable, this excludes exotic quarks $J$ and heavy fields $(U,V)$ although they are $P$-nontrivial too. Because the matter parity $P$ is conserved throughout all the phases of symmetry breaking, it may be applied for low energy processes examined below. Additionally, we are investigating the most interesting setup according to $q=0$, thus $\beta=-1/\sqrt{3}$ and $Y=-1/\sqrt{3}T_8+X$, since this kind of the model reveals a neutrino mass generation scheme at TeV, different from the $U(1)_M$ physics at much large scale \cite{d13,mp2,d17}; the other model according to $q\neq 0$ does not supply such a TeV scheme for neutrino mass. The choice $q=0$ also supplies a variety of types for dark matter, unlike the value $q=-1$ and even no candidate for $q\neq 0,-1$.\footnote{For the last case with $q\neq 0,-1$, which especially includes the minimal 3-3-1 model with $q=1$, one may introduce a vectorlike fermion $F$ that is a 3-3-1 singlet and has $B-L=2k/3$ for $k$ integer. This field is $P$-odd, providing a dark matter candidate. It is noted that the three right-handed neutrinos $\nu_R$'s necessarily for $B-L$ anomaly cancellation are $P$-even, not appropriate to be dark matter.} Also, we consider only $n=0$, thus $P=-1$ for isopartners, without loss of generality.  

\section{\label{fdphase} $U(1)_{T_8}\otimes U(1)_X$ model: family-dependent abelian phase}

In the standard model, the hypercharge $Y$ is family universal; additionally, its mixed anomaly with weak isospin vanishes for each family, i.e. $[SU(2)_L]^2 U(1)_Y\sim 3 Y_{q_L}+Y_{l_L}=0$, where $3$ is the color number. On this theoretical ground, the family number may be left arbitrarily, which is opposite to observation; additionally, the matching $3=-Y_{l_L}/Y_{q_L}$ is questionable, because the color charge and the hypercharge have a distinct nature. Interestingly enough, the abelian charges $T_8$ and $X$ do not obey the properties of $Y$, since the anomaly $[SU(2)_L]^2 U(1)_D \sim 3 D_{q_L}+D_{l_L}\neq 0$ for each family and $D$ is not family universal, for $D=T_8,X$. Even not referring to 3-3-1 symmetry, the $T_8,X$ anomaly cancelation constrains the number of families to be that of colors, similar to a model studied in \cite{dongfc}. Although $T_8$ and $X$ have a nature distinct from $Y$, their combination $Y=\beta T_8+X$ yields the matching and the universality for $Y$. In other words, the standard model with three observed families may be induced by a family-dependent abelian phase. In the following, we will specify the particle content and total Lagrangian of the $U(1)_{T_8}\otimes U(1)_X$ model. We then diagonalize the gauge boson sector, determining the $W$-mass deviation and FCNC. 

\subsection{Particle content and Lagrangian}

\begin{table}[h]
\bc
\begin{tabular}{lccccc}
\hline\hline 
Field & $P$ & $SU(3)_C$ & $SU(2)_L$ & $U(1)_{T_8}$ & $U(1)_X$ \\
\hline 
$l_{aL}=(
\nu_{aL},
e_{aL})$ & $+$ & 1 & 2 & $\fr{1}{2\sqrt{3}}$ & $-\fr 1 3$  \\
$N_{aL}$ & $-$ & 1 & 1 & $-\fr{1}{\sqrt{3}}$ & $-\fr 1 3$\\
$q_{\al L}=(
u_{\al L},
d_{\al L})$ & $+$ & 3 & 2 & $-\fr{1}{2\sqrt{3}}$ & $0$  \\
$J_{\al L}$ & $-$ & 3 & 1 & $\fr{1}{\sqrt{3}}$ & $0$ \\
$q_{3 L}=(
u_{3 L},
d_{3 L})$ & $+$ & 3 & 2 & $\fr{1}{2\sqrt{3}}$ & $\fr 1 3$ \\
$J_{3 L}$ & $-$ & 3 & 1 & $-\fr{1}{\sqrt{3}}$ & $\fr 1 3$ \\
$e_{aR}$ & $+$ & 1 & 1 & $0$ & $-1$ \\
$u_{a R}$ & $+$ & 3 & 1 & $0$ & $\fr 2 3$\\
$d_{a R}$ & $+$ & 3 & 1 & $0$ & $-\fr 1 3$\\
$N_{aR}$ & $-$ & 1 & 1 & 0 & 0 \\
$J_{\al R}$ & $-$ & 3 & 1 & 0 & $-\fr 1 3$ \\
$J_{3 R}$ & $-$ & 3 & 1 & 0 & $\fr 2 3$ \\
$\eta=(
\eta^0_1,
\eta^-_2)$ & $+$ & 1 & 2 & $\fr{1}{2\sqrt{3}}$ & $-\fr 1 3 $ \\
$\eta^0_3$ & $-$ & 1 & 1 & $-\fr{1}{\sqrt{3}}$ & $-\fr 1 3 $ \\
$\rho=(
\rho^+_1,
\rho^0_2)$ & $+$ & 1 & 2 & $\fr{1}{2\sqrt{3}}$ & $\fr 2 3 $ \\
$\rho^+_3$ & $-$ & 1 & 1 & $-\fr{1}{\sqrt{3}}$ & $\fr 2 3 $ \\
$\chi=(
\chi^0_1,
\chi^-_2)$ & $-$ & 1 & 2 & $\fr{1}{2\sqrt{3}}$ & $-\fr 1 3 $ \\
$\chi^0_3$ & $+$ & 1 & 1 & $-\fr{1}{\sqrt{3}}$ & $-\fr 1 3 $ \\
\hline\hline
\end{tabular}
\caption[]{\label{tab1} Field content in family-dependent abelian phase.}
\ec
\end{table}  

In the family-dependent abelian phase, the gauge symmetry and its particle content are summarized in Table \ref{tab1}, along with the matter parity $P$. Each fermion/Higgs isodoublet or isosinglet that possesses $P=1$ has a corresponding isopartner with $P=-1$ due to the mentioned isodoublet-isosinglet splitting. For convenience, we relabel $\eta\equiv (\eta_1,\eta_2)$ and $\rho\equiv (\rho_1,\rho_2)$ for the normal Higgs isodoublets and $\chi\equiv (\chi_1,\chi_2)$ for the isopartner of $\chi_3$, without confusion with those from the higher symmetry. The isopartners $N$, $\eta_3$, and $\chi_1$ are now electrically neutral, providing a dark matter candidate, whereas $\rho_3$ and $\chi_2$ are electrically charged, unavailable for dark matter. It is clear that six fermion doublets, $l_{aL}$ and $q_{3L}$, have a $T_8$ charge opposite to that of other six fermion doublets, $q_{\al L}$, where quark colors must be counted. Hence, the anomaly $[SU(2)_L]^2U(1)_{T_8}$ is cancelled over three families that match the color number, despite the fact that each family is anomalous.  Since $\mathrm{Tr}(T_8)=0$ for left-handed quarks (including exotic quarks) as well as for left-handed leptons (including exotic leptons) in each family, the anomalies $[SU(3)_C]^2 U(1)_{T_8}$ and $[\mathrm{gravity}]^2 U(1)_{T_8}$ vanish. The anomalies $[SU(2)_L]^2U(1)_{X}$, $[SU(3)_C]^2 U(1)_X$, and $[\mathrm{gravity}]^2 U(1)_{X}$ vanish as a result of substituting $X=1/\sqrt{3}T_8+Y$, for which the relevant anomalies for $T_8$ and $Y$ are obviously cancelled. Finally, it is easily verified that the anomalies $[U(1)_{T_8}]^2 U(1)_X$, $[U(1)_X]^2 U(1)_{T_8}$, $[U(1)_{T_8}]^3$, and $[U(1)_X]^3$ all vanish too. 

Up to gauge fixing and ghost terms, the total Lagrangian takes the form, 
\be \mathcal{L}=\mathcal{L}_{\mathrm{kin}}+\mathcal{L}_{\mathrm{Yuk}}-V,\ee where the kinetic term is 
\bea \mathcal{L}_{\mathrm{kin}} &=&\sum_\Psi \bar{\Psi} i\ga^\mu D_\mu \Psi + \sum_\Phi (D^\mu \Phi)^\dagger (D_\mu \Phi)\crn
&&-\fr 1 4 G_{m\mu\nu} G^{\mu\nu}_m -\fr 1 4 A_{i\mu\nu}A^{\mu\nu}_i -\fr 1 4 A_{8\mu \nu} A_8^{\mu\nu} -\fr 1 4 B_{\mu\nu} B^{\mu\nu},\eea where $\Psi$ and $\Phi$ are summed over every fermion and scalar multiplets, respectively, and the covariant derivative and field strengths are  
\bea && D_\mu = \pa_\mu + i g_s t_m G_{m\mu} + i g T_i A_{i\mu}+ i g_1 T_8 A_{8\mu} + i g_2 X B_\mu,\crn
&& G_{m\mu\nu}=\pa_\mu G_{m\nu}-\pa_\nu G_{m \mu} - g_s f_{mnp} G_{n\mu}G_{p\nu},\crn
&& A_{i\mu\nu}=\pa_\mu A_{i\nu}-\pa_\nu A_{i \mu} - g \ep_{ijk} A_{j\mu}A_{k\nu},\crn
&& A_{8\mu\nu}=\pa_\mu A_{8\nu} - \pa_\nu A_{8\mu},\hs B_{\mu\nu}=\pa_\mu B_\nu -\pa_\nu B_\mu,\nn \eea where $(g_s,g,g_1, g_2)$, $(t_m, T_i, T_8, X)$, and $(G_{m\mu}, A_{i\mu}, A_{8\mu}, B_\mu)$ correspond to coupling constants, generators, and gauge bosons of $(SU(3)_C, SU(2)_L, U(1)_{T_8}, U(1)_X)$, respectively. The kinetic mixing term between $U(1)_{T_8,X}$ gauge fields is small and suppressed. Indeed, at high energy such term $\mathcal{L}_{\mathrm{kin}}\supset -\fr{\ep}{2}A_{8\mu\nu}B^{\mu\nu}$ is manifestly prevented by the 3-3-1 gauge symmetry. At low energy, $\ep$ is radiatively generated due to higher weak-isospin splitting, which can be obtained by generalizing the result in \cite{km2022} to be $\ep=\fr{g_1g_2}{12\pi^2}\sum_f T_8 (f_L) X(f_L)\ln\fr{m_r}{m_f}$, where $f$ runs over every fermion of complete $SU(3)_L$ multiplets, and $m_r$ is a renormalization scale. Hence, we have \be \ep=\fr{g_1g_2}{72\sqrt{3}\pi^2}\left(\ln\fr{m^2_{J_3}}{m_b m_t}-\ln \fr{m^2_{N_{1L}}}{m_{\nu_e}m_e}-\ln\fr{m^2_{N_{2L}}}{m_{\nu_\mu}m_{\mu}}-\ln\fr{m^2_{N_{3L}}}{m_{\nu_{\tau}}m_\tau}\right)\sim 10^{-4},\ee which is radically smaller than the relevant mass mixing discussed below. The kinetic mixing may also come from a GUT effect through effective interactions $\fr{c_1}{\La_{\mathrm{GUT}}}S_m A_{m\mu\nu}B^{\mu\nu}+\fr{c_2}{\La^2_{\mathrm{GUT}}}\chi^\dagger T_m \chi A_{m\mu\nu} B^{\mu\nu}+\fr{c_3}{\La^2_{\mathrm{GUT}}}\eta^\dagger T_m \eta A_{m\mu\nu} B^{\mu\nu}+\fr{c_4}{\La^2_{\mathrm{GUT}}}\rho^\dagger T_m \rho A_{m\mu\nu} B^{\mu\nu}$, as coupled to the scalar octet $S$ and triplets $\chi,\eta,\rho$, respectively. The octet dominantly contributes to the mixing, such as $\ep_{\mathrm{GUT}}\sim \fr{\La}{\La_{\mathrm{GUT}}}\sim 10^{-11}$, for $\La_{\mathrm{GUT}}\sim 10^{16}$ GeV and $\La\sim 100$ TeV (see below). That said, the GUT effect is much smaller than the higher weak-isospin splitting.

The Yukawa term and the scalar potential are given, respectively, by 
\bea \mathcal{L}_{\mathrm{Yuk}} &=& h^d_{\al b} \bar{q}_{\al L} \tilde{\eta} d_{bR}+h^u_{3b}\bar{q}_{3L} \eta u_{bR}+x^d_{\al b}\bar{J}_{\al L} \eta^*_3 d_{bR}+x^u_{3b}\bar{J}_{3L}\eta_3 u_{bR}\crn
&& + h^u_{\al b} \bar{q}_{\al L} \tilde{\rho} u_{bR}+h^d_{3b}\bar{q}_{3L} \rho d_{bR}+x^u_{\al b}\bar{J}_{\al L} \rho^{-}_3 u_{bR}+x^d_{3b}\bar{J}_{3L}\rho^+_3 d_{bR}  \crn
&&+ h^J_{\al \beta } \bar{J}_{\al L}\chi^*_3 J_{\beta R} + h^J_{33} \bar{J}_{3L}\chi_3 J_{3R} + y^J_{\al \beta} \bar{q}_{\al L} \tilde{\chi} J_{\beta R} +y^J_{33} \bar{q}_{3L} \chi J_{3R}\crn
&&+h^e_{ab}\bar{l}_{aL}\rho e_{bR}+z_{ab} \bar{N}_{aL} \rho^+_3 e_{bR} +h^N_{ab} \bar{N}_{aL} \chi_3 N_{bR} +t_{ab}\bar{l}_{aL} \chi N_{bR}\crn
&& - \fr 1 2 \mu_{ab} \bar{N}^c_{aR} N_{bR}+H.c.\label{yuk1}
\eea
\bea 
V &=& \mu^2_1 \eta^\dagger \eta + \mu^2_2 \rho^\dagger \rho + \bar{\mu}^2_3 \chi^\dagger \chi+\bar{\mu}^2_1 \eta^*_3 \eta_3 +\bar{\mu}^2_2 \rho_3^- \rho_3^+ + \mu^2_3 \chi^*_3 \chi_3 \crn
&& + \la_1(\eta^\dagger \eta)^2+\la_2(\rho^\dagger \rho)^2+\bar{\la}_3 (\chi^\dagger \chi)^2+ \bar{\la}_1 (\eta^*_3 \eta_3)^2 +\bar{\la}_2 (\rho_3^- \rho_3^+)^2 + \la_3 (\chi^*_3 \chi_3)^2\crn
&& +\la_4 (\eta^\dagger \eta)(\rho^\dagger \rho) +\la_5 (\eta^\dagger \eta) (\chi^\dagger \chi) + \la_6 (\rho^\dagger \rho)(\chi^\dagger \chi) + \la_{7}(\eta^\dagger \rho) (\rho^\dagger \eta)\crn
&&+\la_{8}(\eta^\dagger \chi) (\chi^\dagger \eta)+\la_{9}(\rho^\dagger \chi) (\chi^\dagger \rho)+\fr 1 2 \la_{10}[(\eta^\dagger \chi)^2+H.c.]+\la_{11} (\eta^*_3 \eta_3)(\rho_3^- \rho_3^+) \crn
&& +\la_{12} (\eta^*_3 \eta_3)(\chi_3^* \chi_3 )+\la_{13} (\rho_3^- \rho_3^+)(\chi^*_3 \chi_3)+\fr 1 2 \la_{14}[(\eta^*_3\chi_3)^2+H.c.]\crn
&&+\eta^*_3\eta_3 (\la_{15} \eta^\dagger \eta +\la_{16}\rho^\dagger \rho +\la_{17} \chi^\dagger \chi)+\rho^+_3\rho^-_3 (\la_{18} \eta^\dagger \eta +\la_{19}\rho^\dagger \rho +\la_{20} \chi^\dagger \chi)\crn
&&+ \chi^*_3\chi_3 (\la_{21} \eta^\dagger \eta +\la_{22}\rho^\dagger \rho +\la_{23} \chi^\dagger \chi)+[\eta^*_3\chi_3(\la_{24}\eta^\dagger \chi+\la_{25}\chi^\dagger \eta)+H.c.].\label{potent1}\eea    
Above, $h$'s, $x$'s, $y$'s, $z$'s, $t$'s, and $\la$'s are dimensionless, while $\mu$'s have a mass dimension, and we define $\tilde{X}=i\sigma_2 X^*$ for $X=\eta,\rho,\chi$. Particularly, to make sure that  (i) the scalar fields have a vacuum structure properly (shown below) and (ii) the potential is bounded from below (vacuum stability), the potential parameters satisfy
\be \mu^2_{1,2,3}<0,\hs \bar{\mu}^2_{1,2,3}>0,\hs \la_{1,2,3}>0,\hs \bar{\la}_{1,2,3}>0.\ee The conditions for scalar self-couplings are obtained by requiring $V>0$ when the relevant scalar fields separately tend to infinity. The vacuum stability also demands that $V>0$ when two or more scalar fields simultaneously tend to infinity, which would supply many other conditions for scalar self-couplings, which will be skipped for simplicity.  

The gauge symmetry breaking goes through two stages, 
\bc \begin{tabular}{c} 
$SU(3)_C\otimes SU(2)_L \otimes U(1)_{T_8}\otimes U(1)_X$ \\
$\downarrow w$\\
$SU(3)_C\otimes SU(2)_L \otimes U(1)_Y$ \\
$\downarrow v_{1,2}$\\
$SU(3)_C\otimes U(1)_Q$ \\
\end{tabular}\ec The first stage induced by $\langle \chi^0_3\rangle =w/\sqrt{2} $ has been given, while the last stage is done by  
\be 
\langle \eta \rangle = \begin{pmatrix} \fr{v_1}{\sqrt{2}}\\ 0\end{pmatrix},\hs \langle \rho \rangle =\begin{pmatrix}0\\ \fr{v_2}{\sqrt{2}}\end{pmatrix}.\ee Note that $\chi^0_1$ and $\eta^0_3$ have vanished VEVs, prevented by the matter parity. We demand $w\gg v_{1,2}$ and $v^2_1+v^2_2=(246\ \mathrm{GeV})^2$ for consistency with the standard model.

\subsection{Gauge sector and $W$-mass deviation}

Substituting the VEVs of scalar fields to the kinetic term $\mathcal{L}\supset \sum_\Phi (D^\mu \Phi)^\dagger (D_\mu \Phi)$, gauge boson masses are given. First note that gluons are massless since the color charge is unbroken. For electroweak sector, the charged gauge boson $W^\pm=(A_1\mp i A_2)/\sqrt{2}$ is a physical field, i.e. a mass eigenstate, by itself with mass, \be m^2_W=\fr{g^2}{4}(v^2_1+v^2_2).\ee By contrast, the neutral gauge fields $A_3$, $A_8$, and $B$ mix through a mass matrix, \be \mathcal{L}\supset \fr 1 2 \begin{pmatrix} A_3 & A_8 & B\end{pmatrix} M^2 \begin{pmatrix} A_3\\ A_8\\ B\end{pmatrix},\ee where   
\bea M^2=\fr{g^2}{4}\begin{pmatrix} v^2_1+v^2_2 & \fr{t_1}{\sqrt{3}}(v^2_1-v^2_2)& -\fr{2t_2}{3}(v^2_1+2v^2_2)\\
\fr{t_1}{\sqrt{3}}(v^2_1-v^2_2) & \fr{t^2_1}{3}(v^2_1+v_2^2+4w^2) & -\fr{2t_1 t_2}{3\sqrt{3}}(v^2_1-2v^2_2-2w^2)\\
-\fr{2t_2}{3}(v^2_1+2v^2_2) & -\fr{2t_1 t_2}{3\sqrt{3}}(v^2_1-2v^2_2-2w^2) & \fr{4t^2_2}{9}(v^2_1+4v^2_2+w^2)
\end{pmatrix}, \eea where we have defined $t_1=g_1/g$ and $t_2=g_2/g$.

The mass matrix $M^2$ has an exact zero eigenvalue corresponding to the photon mass with an exact mass eigenstate corresponding to the photon field as 
\be \fr{A}{e}=\fr{A_3}{g}-\fr{A_8}{\sqrt{3}g_1}+\fr{B}{g_2},\ee normalized by \be \fr{1}{e^2}=\fr{1}{g^2}+\fr{1}{3g^2_1}+\fr{1}{g^2_2}.\ee Alternatively, since the photon field couples to the electric charge, $Q=T_3+Y=T_3-T_8/\sqrt{3}+X$, the above solution can be derived by replacing every generator in the electric charge relation by corresponding gauge field over coupling, i.e. $Q$ by $A/e$, $T_3$ by $A_3/g$, $T_8$ by $A_8/g_1$, and $X$ by $B/g_2$. Further, the electromagnetic coupling is related to the sine of the Weinberg's angle as $e=g s_W$, while the tan of this angle is $t_W=g_Y/g$, where the hypercharge coupling is given by $g_Y=\sqrt{3}g_1 g_2/\sqrt{3g_1^2+g^2_2}$. Additionally, we define $g_Y=-\sqrt{3}g_1 s_\theta = g_2 c_\theta$, i.e. $t_\theta=-g_2/\sqrt{3}g_1$. The photon field is rewritten as 
\be A=s_W A_3+ c_W (s_\theta A_8 + c_\theta B).\ee The standard model $Z$ field is given orthogonally to $A$, 
\be Z=c_W A_3 - s_W (s_\theta A_8 + c_\theta B),\ee as usual, while a new field, 
\be Z'= c_\theta A_8 - s_\theta B,\label{zprime1}\ee is identified, orthogonal to the hypercharge field in parentheses in $A,Z$.   

In the new basis with $(A,Z,Z')$, the photon field is decoupled as physical field, while the rest $Z$ and $Z'$ mix through a mass matrix, \be \mathcal{L}\supset \fr 1 2 \begin{pmatrix}Z & Z'\end{pmatrix} \begin{pmatrix} m^2_Z & m^2_{ZZ'} \\
m^2_{ZZ'} & m^2_{Z'}\end{pmatrix}\begin{pmatrix} Z\\ Z'\end{pmatrix},\ee where  
\bea m^2_Z &=& \fr{g^2}{4 c^2_W}(v^2_1+v^2_2),\\
m^2_{ZZ'}&=&\fr{g^2t_W}{6s_{2\theta}c_W}\left[(1+3s^2_\theta)v^2_2-(1-3s^2_\theta)v^2_1\right],\\
m^2_{Z'}&=&\fr{g^2 t^2_W}{9s^2_{2\theta}}\left[(1+3s^2_\theta)^2v^2_2+(1-3s^2_\theta)^2v^2_1+4w^2\right].\eea Diagonalizing this mass matrix, we achieve two physical fields,
\be Z_1= c_\varphi Z - s_\varphi Z',\hs Z_2 = s_\varphi Z + c_\varphi Z',\ee where the $Z$-$Z'$ mixing angle $(\varphi)$ is defined by \be t_{2\varphi}=\fr{2m^2_{ZZ'}}{m^2_{Z'}-m^2_Z}\simeq \fr 3 4 \fr{s_{2\theta}}{s_W}\fr{\left[(1+3s^2_\theta)v^2_2-(1-3s^2_\theta)v^2_1\right]}{w^2},\ee and the masses of $Z_1$ and $Z_2$ satisfy
\bea m^2_{Z_1} &=& \fr 1 2 \left[m^2_Z + m^2_{Z'}-\sqrt{(m^2_Z-m^2_{Z'})^2+ 4 m^4_{ZZ'}}\right]\simeq m^2_Z-\fr{m^4_{ZZ'}}{m^2_{Z'}},\\ 
m^2_{Z_2} &=& \fr 1 2 \left[m^2_Z + m^2_{Z'}+\sqrt{(m^2_Z-m^2_{Z'})^2+ 4 m^4_{ZZ'}}\right]\simeq m^2_{Z'}.\eea The $Z$-$Z'$ mixing angle is small, suppressed by $(v_1,v_2)^2/w^2$. Additionally, the field $Z_1$ possesses a mass approximating that of $Z$, called the standard model $Z$-like boson, while the field $Z_2$ is a new heavy gauge field with mass at $w$ scale. 

Because of the $Z$-$Z'$ mixing, the observed $Z_1$ mass is reduced in comparison with the standard model $Z$ mass. Because of this mass reduction, it gives rise to a positive contribution to the $T$-parameter, such as 
\be \al T = \rho-1=\fr{m^2_W}{c^2_W m^2_{Z_1}}-1\simeq \fr{m^4_{ZZ'}}{m^2_Z m^2_{Z'}}\simeq \fr{\left[(1+3s^2_\theta)v^2_2-(1-3s^2_\theta)v^2_1\right]^2}{4(v^2_1+v^2_2)w^2}.\ee Since the $Z_1$ mass is precisely measured, the positive value $\al T$ mainly enhances the $W$-boson mass, such as \cite{stu,strumia}
\be \Delta m^2_W=\fr{c^4_W m^2_Z}{c^2_W-s^2_W}\al T \simeq \fr{c^2_W g^2 \left[(1+3s^2_\theta)v^2_2-(1-3s^2_\theta)v^2_1\right]^2}{16c_{2W}w^2}.\ee

\begin{figure}[h]
\bc
\includegraphics[scale=0.64]{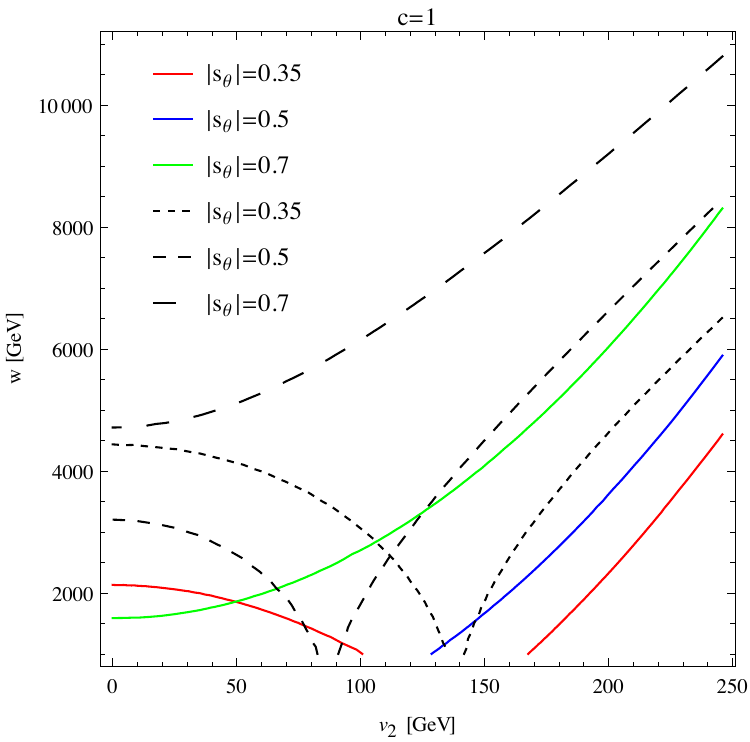}
\includegraphics[scale=0.63]{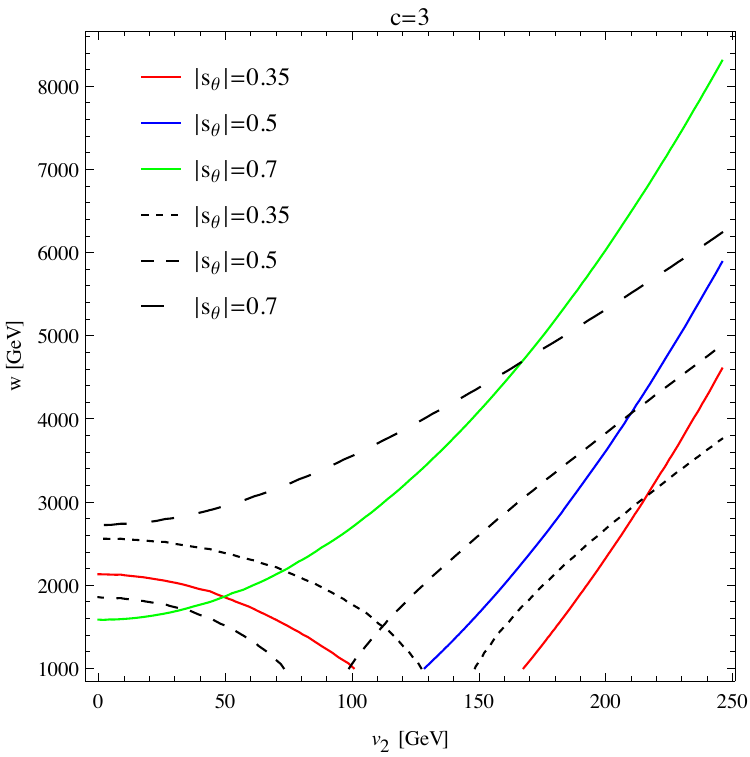}
\caption[]{\label{fig1} CDF $W$-mass deviation (solid lines) and EWPT $Z$-$Z'$ mixing angle $|\varphi|=c\times 10^{-3}$ for $c=1$ and 3 (dashed lines) contoured as functions of $(v_2,w)$ for fixed values of $|s_\theta|$.}
\ec
\end{figure}

The standard model predicts $m_W|_\mathrm{SM}=80.357 \pm 0.006$ GeV, which deviates from the CDF measurement $m_W|_{\mathrm{CDF}}=80.4335\pm 0.0094$ GeV at 7$\sigma$ \cite{CDF:2022hxs}. Additionally, the precision measurements of $Z$-pole physics at the LEP and SLC make a strong constraint on the $Z$-$Z'$ mixing angle $\varphi$ to be of the order of $10^{-3}$ \cite{zzprimemixing}. Hence, we contour both $\Delta m^2_W=m^2_W|_{\mathrm{CDF}}-m^2_W|_{\mathrm{SM}}$ taking the difference of centered values and $|\varphi|=c\times 10^{-3}$ concretely inputing $c=1$ and $3$ as functions of $v_2$ and $w$ in Fig. \ref{fig1}. Here, notice that $v_1=\sqrt{(246\ \mathrm{GeV})^2-v^2_2}$, $s^2_W=0.231$, and $\al=1/128$. Additionally, at $SU(3)_L$ breaking scale $\La$, the couplings $g$ and $g_1$ match, i.e. $g=g_1$, as unified. When the energy decreases from $\La$ to $w$, the $SU(2)_L$ coupling $g$ increases while the $U(1)_{T_8}$ coupling $g_1$ decreases, as generically implied by the RGE. It yields $g>g_1$ at $w$ scale, implying $|s_\theta|=\fr{t_W}{\sqrt{3}}\fr{g}{g_1}>\fr{t_W}{\sqrt{3}}=0.316$. Hence, we have taken $|s_\theta|=0.35$, $0.5$, and $0.7$ for the above contours. Obviously, the $W$-mass deviation is solved if the solid line lies above the dashed line, for each $|s_\theta|$. Hence, given that $c = 3$, we obtain viable parameter regime $(v_2>216,w>3075)$, $(v_2>210,w>4045)$, and $(v_2>167,w>4686)$, all in GeV, according to the choice of $|s_\theta|=0.35$, 0.5, and 0.7, respectively. Otherwise, if $c=1$, or exactly $|\varphi| < 10^{-3}$, the model cannot explain the $W$-boson mass anomaly, unless $|s_\theta|\rightarrow 1$.   

Last, but not least, this study suppresses a kinetic mixing effect between the two $U(1)_{T_8,X}$ gauge fields---which may be experienced in this family-dependent abelian phase but absolutely does not occur in the 3-3-1 phase due to the symmetry protection---which may modify the current result. However, due to the symmetry protection, such effect that arises at low energy would be small, which may be neglected. 

On the other hand, the usual 3-3-1 model also predicts the $W$-mass shift that corresponds to the case $|s_\theta|=0.316$, i.e. $g=g_1$ and $\La$ close to $w$ \cite{wm331}. The present result according to $c=3$ indicates that the $Z$-$Z'$ mixing source responsible for the $W$-mass anomaly in the usual 3-3-1 model would possess a narrow (viable) parameter regime, $v_2\to 246$ GeV, limited by the EWPT $Z$-$Z'$ mixing angle and the FCNC below.        

\subsection{Tree-level FCNC}

To extract FCNC, we consider the fermion couplings with neutral gauge bosons,
\be \mathcal{L}\supset -\sum_\Psi \bar{\Psi}\ga^\mu [g T_3 A_{3\mu}+g_1 T_8 A_{8\mu}+g_2 (T_8/\sqrt{3}+Y) B_\mu]\Psi,\ee where $X=T_8/\sqrt{3}+Y$ that couples to $B$ is substituted. Since exotic fermions $N,J$ are matter parity odd, impossibly mixed with ordinary fermions, the charges $T_3$ and $Y$ conserve all flavors. The Lagrangian that leads to FCNC includes only $T_8$ couplings, 
\be \mathcal{L} \supset -\sum_\Psi \bar{\Psi}\ga^\mu T_8 \Psi [g_1  A_{8\mu}+(g_2/\sqrt{3}) B_\mu].\ee
Since $T_8$ acts only on left chirality and conserves both lepton flavors and exotic quark flavors, the FCNC is only relevant to ordinary left-handed quarks, i.e. $\Psi\to q_L$. Additionally, we have $g_1 A_8+(g_2/\sqrt{3}) B= -(gt_W/\sqrt{3}s_\theta c_\theta)Z'$, achieved from (\ref{zprime1}) with notice that $g_1=-gt_W\sqrt{3}s_\theta$ and $g_2=gt_W/c_\theta$. Hence, the Lagrangian becomes
\be \mathcal{L} \supset \fr{gt_W}{\sqrt{3}s_\theta c_\theta}\sum_{q_L} \bar{q}_L\ga^\mu T_8 q_L Z', \ee  Further, because of $T_8(q_{1,2L})=-1/2\sqrt{3}=-T_8(q_{3L})$ and by substituting mass eigenstates of quarks, i.e. $q_{aL}=[V_{qL}]_{ai}q_{iL}$ with $V^\dagger_{qL}V_{qL}=1$, we obtain the FCNC, 
 \be \mathcal{L} \supset \fr{gt_W}{3s_\theta c_\theta} \sum_{i,j}[V^*_{qL}]_{3i} [V_{qL}]_{3j}\bar{q}_{iL}\ga^\mu q_{jL} Z'_\mu,\ee which flavor changes for $i\neq j$, where $q$ denotes either up-type or down-type quarks. 

Integrating $Z'$ out, we achieve an effective Hamiltonian describing neutral meson mixing,
\be \mathcal{H}_{\mathrm{eff}}=\fr{([V^*_{qL}]_{3i} [V_{qL}]_{3j})^2}{w^2} (\bar{q}_{iL}\ga^\mu q_{jL})^2,\label{ptdt11}\ee where we have used $m^2_{Z'}\simeq (g^2 t^2_W/9s^2_\theta c^2_\theta)w^2$. Notice that due to $Z$-$Z'$ mixing, the $Z$ boson also contributes to the effective Hamiltonian but being more suppressed than $Z'$, since $t_\varphi \sim m^2_Z/m^2_{Z'}\ll m_Z/m_{Z'}$. The effective Hamiltonian induced is independent of $v_{1,2}$ and $\theta$ due to $w\gg v_{1,2}$. Additionally, it does not depend on the new physics phase associated with $\La$ even for $\La\gg w$ or $\La\sim w$ or even $\La$ not existed, since $\La$ does not contribute to neutral gauge boson masses. In other words, the above result applies for every 3-3-1 model, including the family-dependent abelian model. 

The effective interactions (\ref{ptdt11}) contribute to neutral-meson mixing amplitudes, which have been extensively studied. The existing data requires 
\bea && \fr{([V^*_{dL}]_{31} [V_{dL}]_{32})^2}{w^2}<\left(\fr{1}{10^4\ \mathrm{TeV}}\right)^2,\\ &&\fr{([V^*_{dL}]_{31} [V_{dL}]_{33})^2}{w^2}<\left(\fr{1}{500\ \mathrm{TeV}}\right)^2,\\ 
&&\fr{([V^*_{dL}]_{32} [V_{dL}]_{33})^2}{w^2}<\left(\fr{1}{100\ \mathrm{TeV}}\right)^2,\label{ptdt15}\eea corresponding to $K^0$-$\bar{K}^0$ mixing ($ij=12$), $B^0_d$-$\bar{B}^0_d$ mixing ($ij=13$), and $B^0_s$-$\bar{B}^0_s$ mixing ($ij=23$), respectively \cite{fcnceff,fcnceff1}. Particularly aligning quark mixing to the down sector, we have $V_{dL}=V_{\mathrm{CKM}}$, which is well measured, such as $[V_{dL}]_{31}=0.00886$, $[V_{dL}]_{32}=0.0405$, and $[V_{dL}]_{33}=0.99914$ \cite{pdg}. Hence, the above constraints yield
\be w> 3.6\ \mathrm{TeV},\hs w>4.4\ \mathrm{TeV},\hs w>4\ \mathrm{TeV},\label{ptdt112}\ee corresponding to the mentioned meson mixings, which are in good agreement with the CDF $W$-mass and EWPT $Z$-$Z'$ mixing bounds.        

\section{\label{fmass} Neutrino mass generation: Scotogenic setup}

The Yukawa Lagrangian in (\ref{yuk1}) will generate tree-level masses for all relevant fermions, when the scalar fields develop VEVs. The isopartners $\chi^0_1$ and $\eta^0_3$ have vanished VEV due to the matter parity conservation. However, $\eta^0_1$, $\rho^0_2$, and $\chi^0_3$ can develop the VEVs, as given, inducing tree-level fermion masses. It is noted that since the matter parity is conserved, usual quarks and exotic quarks, as well as usual leptons and new leptons, do not mix. We first give a comment on charged fermions. The masses of exotic quarks are \be [m_J]_{\al\beta}=-h^J_{\al\beta}\fr{w}{\sqrt{2}},\hs [m_J]_{33}=-h^J_{33}\fr{w}{\sqrt{2}},\ee which are large at $w$ scale as appropriate. The masses of up-type quarks are
\be [m_u]_{\al b}=-h^u_{\al b}\fr{v_2}{\sqrt{2}},\hs [m_u]_{3b}=-h^u_{3b}\fr{v_1}{\sqrt{2}},\ee while those of down-type quarks are 
\be [m_d]_{\al b}=h^d_{\al b}\fr{v_1}{\sqrt{2}},\hs [m_d]_{3b}=-h^d_{3b}\fr{v_2}{\sqrt{2}},\ee which are all at weak scales $v_{1,2}$. Diagonalizing these matrices, we get consistent masses for $u$, $d$, $c$, $s$, $t$, and $b$ as well as CKM matrix. Similarly, charged leptons gain suitable masses, \be [m_e]_{ab}=-h^e_{ab}\fr{v_2}{\sqrt{2}},\ee analogous to the standard model.

The Lagrangian that generates tree-level masses for neutral fermions contains,
\be \mathcal{L}\supset t_{ab}\bar{l}_{aL} \chi N_{bR} +h^N_{ab} \bar{N}_{aL} \chi_3 N_{bR}   - \fr 1 2 \mu_{ab} \bar{N}^c_{aR} N_{bR}+H.c.\label{ma1} \ee  The last two terms supply new lepton mass as
\be \mathcal{L}\supset -\fr 1 2 \begin{pmatrix} \bar{N}_L& \bar{N}^c_R \end{pmatrix}\begin{pmatrix}0 & m\\
m^T & \mu \end{pmatrix}\begin{pmatrix} N^c_L \\
N_R \end{pmatrix}+H.c.,\label{nlss}\ee where $m=-h^N w/\sqrt{2}$, and family indices are suppressed for brevity. The new lepton $N$ has a Dirac mass $m$ naturally given at $w$ scale. Whereas, $N_R$ possesses a bare Majorana mass $\mu$, not protected by symmetry, since it is a singlet and has $B-L=0$. Hence, $\mu$ may be arbitrarily large, which is naturally taken at $\La$ scale. In other words, $\mu\gg w$ is appropriately imposed. The above mass matrix obeys a seesaw form, such that 
\be m_{N_L}\simeq -m\mu^{-1} m^T\sim w^2/\mu,\hs m_{N_R}\simeq \mu, \ee while the $N_L$-$N_R$ mixing is small and neglected. Since $w\sim$ TeV, we take $\mu\sim \La \sim 100$ TeV, as expected. Therefore, we obtain $m_{N_L}\sim 10$ GeV, while $m_{N_R}\sim 100$ TeV.

The first term in (\ref{ma1}) couples lepton doublets to the isopartners which are matter parity odd, an inert scalar doublet $\chi$ and Majorana fermion singlets $N_R$'s. This fully realizes a scotogenic scheme for neutrino mass generation \cite{sgsetup}. Indeed, the field $\chi$ couples to normal Higgs doublets and singlet in the potential (\ref{potent1}), collected as   
\bea V &\supset& \bar{\mu}^2_3 \chi^\dagger \chi+\la_5 (\eta^\dagger \eta) (\chi^\dagger \chi) + \la_6 (\rho^\dagger \rho)(\chi^\dagger \chi) +\la_{8}(\eta^\dagger \chi) (\chi^\dagger \eta)\crn
&&+\la_{9}(\rho^\dagger \chi) (\chi^\dagger \rho)+\fr 1 2 \la_{10}[(\eta^\dagger \chi)^2+H.c.]+\la_{23} (\chi^*_3\chi_3)(\chi^\dagger \chi), \eea where the bare mass term of $\chi$, i.e. $\bar{\mu}_3$, is concerned as also included. When $\eta,\rho,\chi_3$ develop the VEVs $v_{1},v_{2},w$ respectively, the field \be \chi=\begin{pmatrix}\chi^0_1\\ \chi^-_2\end{pmatrix}=\begin{pmatrix} \fr{1}{\sqrt{2}}(R_1+ i I_1)\\ \chi^-_2\end{pmatrix}\ee is separated in mass, such as
\bea m^2_{R_1}&=&\bar{\mu}^2_3+\fr{\la_5+\la_8+\la_{10}}{2}v^2_1+\fr{\la_6}{2}v_2^2+\fr{\la_{23}}{2}w^2,\\
m^2_{I_1}&=&\bar{\mu}^2_3+\fr{\la_5+\la_8-\la_{10}}{2}v^2_1+\fr{\la_6}{2}v_2^2+\fr{\la_{23}}{2}w^2,\\
m^2_{\chi_2}&=&\bar{\mu}^2_3+\fr{\la_5}{2}v^2_1+\fr{\la_6+\la_9}{2}v_2^2+\fr{\la_{23}}{2}w^2. \eea  There exist $R_1$-$\Re(\eta_3)$ mixing and $I_1$-$\Im(\eta_3)$ mixing due to $\la_{24,25}$ couplings. However, the mixing effects are small, suppressed by $(v_1/w)$, which can be neglected. The physical fields $R_1$, $I_1$, and $\chi_2$ obtain a mass naturally given at $w$ scale. Especially, the $R_1$ and $I_1$ masses are small separated by $\la_{10}$ coupling, such as \be m^2_{R_1}-m^2_{I_1}=\la_{10}v^2_1\ll m^2_{\chi_1},\label{mspx1}\ee where $m^2_{\chi_1}\equiv (m^2_{R_1}+m^2_{I_1})/2\simeq \bar{\mu}^2_3+(\la_{23}/2)w^2$. 

\begin{figure}[h]
\bc
\includegraphics[]{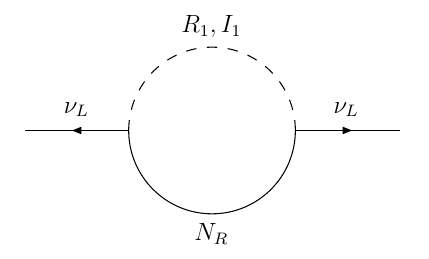}
\caption[]{\label{fig2} Dark isopartner contribution to neutrino mass.}
\ec
\end{figure}

That said, the neutrino generation Lagrangian is given by 
\be \mathcal{L}\supset \fr{t_{ab}}{\sqrt{2}}\bar{\nu}_{aL}(R_1+iI_1)N_{bR}-\fr 1 2 m_{N_{bR}} N^2_{bR}-\fr 1 2 m^2_{R_1} R^2_1-\fr 1 2 m^2_{I_1} I^2_1,\ee where we assume the mass matrix $m_{N_R}=\mathrm{diag}(m_{N_{1R}},m_{N_{2R}}, m_{N_{3R}})$ to be flavor diagonal, i.e. $N_{bR}$ is a physical Majorana field by itself, without loss of generality. The neutrino mass is induced by one-loop diagrams mediated by $N_R$ and $R_1/I_1$, as depicted in Fig. \ref{fig2}. Note that in the gauge basis, the $N_R$ line is attached by Majorana mass term $\mu_{ab}$, while the $R_1/I_1$, i.e. $\chi_1$, line is coupled to external $\eta_1$ field via $\la_{10}$ coupling, analogous to that in \cite{sgsetup}. 

The neutrino mass matrix that is derived in form of $\mathcal{L}\supset -\fr 1 2 \bar{\nu}_{aL} [m_\nu]_{ab} \nu^c_{bL}+H.c.$ reads
\bea -i [m_\nu]_{ab}P_R&=&\int\fr{d^4 p }{(2\pi)^4}\left(i\fr{t_{ak}}{\sqrt{2}} P_R\right) \fr{i}{\slash\!\!\! p-m_{N_{kR}}}
\left(i\fr{t_{bk}}{\sqrt{2}}P_R\right)\fr{i}{p^2- m^2_{R_1}}\crn
&&+\int\fr{d^4 p }{(2\pi)^4}\left(-\fr{t_{ak}}{\sqrt{2}} P_R\right) \fr{i}{\slash\!\!\! p-m_{N_{kR}}}
\left(-\fr{t_{bk}}{\sqrt{2}} P_R\right)\fr{i}{p^2- m^2_{I_1}}\crn
&=& \fr{P_R}{2}t_{ak}t_{bk} m_{N_{kR}}\int \fr{d^4p}{(2\pi)^4}\fr{m^2_{R_1}-m^2_{I_1}}{(p^2-m^2_{N_{kR}})(p^2-m^2_{R_1})(p^2-m^2_{I_1})}.\label{cvg}\eea Notice that the contributions of $R_1$ and $I_1$ have opposite signs because the product of two neutrino vertices coupled to the real and imaginary parts of $\chi_1$ changes the sign. Hence, the result depends on $R_1,I_1$ mass splitting for which the divergences according to each diagram are cancelled out, matching the convergent integral (\ref{cvg}). Given that $x\gg y,z$, one has
 \be \int \fr{d^4 p}{(2\pi)^4}\fr{1}{(p^2-x )(p^2-y)(p^2-z)} \simeq \fr{i}{16\pi^2}\fr{y\ln y/x-z \ln z/x}{x (y-z)}.\ee
Applying to this model for $m_{N_R}\gg m_{R_1,I_1}$, it gives rise to
\bea [m_\nu]_{ab} &\simeq& -\fr{1}{32\pi^2}\fr{t_{ak}t_{bk}}{m_{N_{kR}}}\left(m^2_{R_1}\ln \fr{m^2_{R_1}}{m^2_{N_{kR}}}-m^2_{I_1}\ln \fr{m^2_{I_1}}{m^2_{N_{kR}}}\right)\crn
&\simeq& \fr{\la_{10}}{32\pi^2} \fr{t_{ak}t_{bk} v^2_1}{m_{N_{kR}}}\left(\ln \fr{m^2_{N_{kR}}}{m^2_{\chi_1}}-1\right),\eea where the last approximation uses the small mass splitting of $R_1,I_1$ in (\ref{mspx1}). The neutrino mass is naturally small, as suppressed by the loop factor $1/32\pi^2$, the large scale $m_{N_{R}}$, and the $B-L$ violating coupling $\la_{10}$ (notice that $\la_{10}$ is also presented in the usual 3-3-1 model and always conserves the matter parity). Indeed, taking $\la_{10}\sim 10^{-3}$, $t\sim 10^{-2}$, $m_{N_R}\sim 100$ TeV, $m_{\chi_1}\sim$ TeV, and $v_1\sim 100$ GeV, we have $m_\nu\sim 0.1$ eV, as expected. 

Last, but not least, the above mechanism cannot work in the usual 3-3-1 model since $\chi^0_1$ is a Goldstone boson, eaten by the corresponding gauge boson $U^0$. Since the gauge interaction of scalar field that induces a gauge boson mass always conserves CP, the field like $U^0$ cannot be separated in mass. Thus, the contribution of $U^0$ to neutrino mass via a diagram similar to Fig. \ref{fig2} by replacing $\chi_1$ by $U$ and $N_R$ by $N_L$ vanish, which is unlike a model in \cite{vdoublet}. We call the reader's attention to recent scotogenic approaches in the 3-3-1-1 model \cite{3311scoto,3311scoto1,3311scoto2}.   

\section{\label{dms} Dark matter stability: Seesaw setup}

The present model predicts the lightest state of $N_{aL}$, denoted $N_L$, to be the lightest dark isopartner---the unique candidate for dark matter. Indeed, this candidate is naturally stabilized by the matter parity $P$ and the seesaw mechanism in (\ref{nlss}), as given. The other dark isopartners, say $\chi^0_1$, $\chi^-_2$, $\eta^0_3$, $\rho^+_3$, and $J_a$, all have a mass at (i.e., proportional to) $w$ scale, while $N_{aR}$ like the vector doublet $(U,V)$ and the Higgs octet $S$ all possess a mass at $\La$ scale. The seesaw mechanism makes the $N_L$ mass $\sim w^2/\La$ naturally (far) below $w$ scale, i.e. setting $N_L$ to be the lightest dark isopartner. Hence, contrary to the scotogenic setup, the present model predicts the light seesaw partner $N_L$ of the heavy scotogenic field $N_R$ to be dark matter. Additionally, this setup differs from the usual 3-3-1 model for which several candidates for dark matter are viable. 

\begin{figure}[h]
\bc
\includegraphics[]{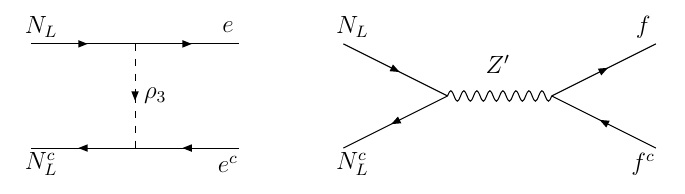}
\caption[]{\label{fig3} Dark matter annihilation to normal matter, where $e$ denotes charged leptons while $f$ indicates usual leptons and quarks (except top quark).}
\ec
\end{figure}                 

The field $N_L$ annihilates to charged leptons via $t$-channel diagrams exchanged by $\rho^\pm_3$ as well as to ordinary fermions (except top quark) via $s$-channel diagrams through $Z'$ portal, as depicted in Fig. \ref{fig3}, which sets $N_L$ relic density. The $t$-channel diagrams are governed the Yukawa couplings $z_{ab}\bar{N}_{aL}\rho^+_3 e_{bR}+H.c.$ as in the Yukawa Lagrangian, while the $t$-channel diagrams are set by the $Z'$ couplings to fermions, such as 
\bea g^{Z'}_V(f) &=&\fr{s_W}{s_\theta c_\theta}\left[s^2_\theta T_3(f_L)-\fr{1}{\sqrt{3}}T_8(f_L)-2s^2_\theta Q(f)\right],\\ 
g^{Z'}_A(f) &=& \fr{s_W}{s_\theta c_\theta}\left[s^2_\theta T_3(f_L) -\fr{1}{\sqrt{3}}T_8(f_L)\right],\eea which are explicitly collected in Tab. \ref{tab3}. Here note that the $Z$-$Z'$ mixing effect is small and neglected.  
\begin{table}[h]
\bc
\begin{tabular}{lcc}
\hline\hline
$f$ & $g^{Z'}_V(f)$ & $g^{Z'}_A(f)$\\
\hline
$\nu_a$ & $\fr{s_W}{s_{2\theta}}\left(-\fr 1 3+s^2_\theta\right)$ & $\fr{s_W}{s_{2\theta}}\left(-\fr 1 3+s^2_\theta\right)$\\
$e_a$ & $\fr{s_W}{s_{2\theta}}\left(-\fr 1 3+3 s^2_\theta\right)$ & $\fr{s_W}{s_{2\theta}}\left(-\fr 1 3-s^2_\theta\right)$ \\
$u_\al $ & $\fr{s_W}{s_{2\theta}}\left(\fr 1 3-\fr 5 3 s^2_\theta\right)$ & $\fr{s_W}{s_{2\theta}}\left(\fr 1 3+s^2_\theta\right)$\\
$u_3$ & $\fr{s_W}{s_{2\theta}}\left(-\fr 1 3-\fr 5 3 s^2_\theta\right)$ & $\fr{s_W}{s_{2\theta}}\left(-\fr 1 3+s^2_\theta\right)$\\
$d_\al $ & $\fr{s_W}{s_{2\theta}}\left(\fr 1 3+\fr 1 3 s^2_\theta\right)$ & $\fr{s_W}{s_{2\theta}}\left(\fr 1 3-s^2_\theta\right)$\\
$d_3$ & $\fr{s_W}{s_{2\theta}}\left(-\fr 1 3+\fr 1 3 s^2_\theta\right)$ & $\fr{s_W}{s_{2\theta}}\left(-\fr 1 3-s^2_\theta\right)$\\
$N_a$ & $\fr{s_W}{3s_\theta c_\theta}$ & $\fr{s_W}{3s_\theta c_\theta}$\\
\hline\hline
\end{tabular}
\caption[]{\label{tab3}Couplings of $Z'$ to fermions.}
\ec
\end{table}
If the $s$-channel diagrams either contribute equivalently to or dominate over the $t$-channel diagrams, this yields an annihilation cross-section proportional to  \be \langle \sigma v\rangle_s \sim \fr{g^4}{16\pi c^4_W}\fr{m^2_{N_L}}{m^4_{Z'}}\simeq \left(\fr{m_{N_L}}{m_{Z'}}\right)^2\times \left(\fr{1.5\ \mathrm{TeV}}{m_{Z'}}\right)^2\times 1\ \mathrm{pb},\ee which is much smaller than the experimental value $\langle \sigma v\rangle\simeq $ 1 pb, because of $m_{N_L}\ll m_{Z'}$. Therefore, the expected annihilation cross-section must be dominated by the $t$-channel diagrams (necessarily enhanced by the $z$ coupling), computed as  
\bea \langle \sigma v\rangle &=& \fr{\sum_{i,j}|z^*_{1 i} z_{1j}|^2}{8\pi} \fr{m^2_{N_L}}{m^4_{\rho_3}}\crn &\simeq& \left(\fr{\sum_{i}|z_{1 i}|^2}{4\pi}\right)^2\times \left(\fr{m_{N_L}}{10\ \mathrm{GeV}}\right)^2\times \left(\fr{700\ \mathrm{GeV}}{m_{\rho_3}}\right)^4\times 1\ \mathrm{pb},\eea where the dark matter $N_L$, assumed as $N_{1L}$ for brevity, is heavier than charged leptons $e_{i/j}$, so the charged lepton masses are omitted. The dark matter relic density $\Om h^2\simeq 0.1\ \mathrm{pb}/\langle \sigma v\rangle \simeq 0.11$ \cite{pdg} is explained, given that $\sum_i |z_{1i}|^2/4\pi \sim 1$, $m_{\rho_3}\sim 700$ GeV, with the dark matter mass $m_{N_L}\sim 10$ GeV as implied by the seesaw mechanism. Although the $Z'$ portal that drives the $s$-channel annihilation does not contribute to the dark matter abundance, it may mark a direct dark matter detection signal, as interpreted hereafter.    

Since $N_L$ is a Majorana field, it only scatters with quarks in direct detection experiment through spin-dependent (SD) effective interaction induced by $Z'$ similar to the right diagram in Fig. \ref{fig3} for $f=q$, such as
\be \mathcal{L}_{\mathrm{eff}}\supset \fr{g^2}{4c^2_Wm^2_{Z'}}[g^{Z'}_A(N)g^{Z'}_A(q)]^2(\bar{N} \ga^\mu \ga_5 N)(\bar{q}\ga_\mu \ga_5q),\ee where $N\equiv N_L+N^c_L$ and the couplings $g^{Z'}_A(f)$ for $f=N,q$ have been given in Tab. \ref{tab3}. We obtain the SD cross-section describing the scatter of $N$ with a target neutron ($n$), 
\be \sigma^{\mathrm{SD}}_{N_L}=\fr{3g^4}{4\pi c^4_W}\fr{m^2_n}{m^4_{Z'}}[g^{Z'}_A(N)]^2[g^{Z'}_A(u)\la^n_u +g^{Z'}_A(d)(\la^n_d+\la^n_s)]^2,\ee where the fractional quark-spin coefficients for neutron are $\la^n_u=-0.42$, $\la^n_d=0.85$, and $\la^n_s=-0.88$ \cite{sddm}. The scatter with proton gives a similar bound and is not considered. 

\begin{figure}[h]
\bc
\includegraphics[]{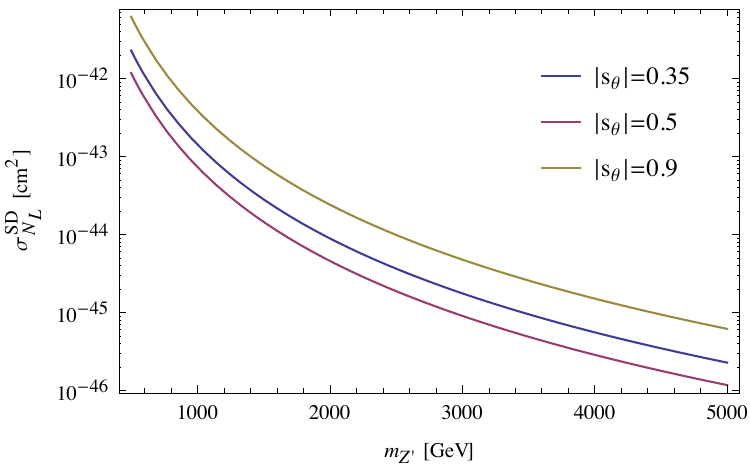}
\caption[]{\label{fig4} SD cross-section of dark matter $N_L$ with neutron plotted as function of $m_{Z'}$ according to the fixed values of $|s_\theta|$.}
\ec
\end{figure}

We take $s^2_W=0.231$, $\al=g^2 s^2_W/4\pi=1/128$, and $m_n=1$ GeV, as usual. In Fig. \ref{fig4}, we plot the SD cross-section as a function of $Z'$ mass according to $|s_\theta|=0.35$, $0.5$, and $0.9$, where the first two values have previously been given, while the last one allows us examining a limit. It is clear that this model predicts the dark matter signal strength in direct detection safely below the current bounds $\sim 10^{-41}\ \mathrm{cm}^2$ for $m_{N_L}\sim 10$ GeV from various experiments~\cite{ddsd}, even for large $|s_\theta|$ and $Z'$ mass beyond the weak scale. The SD cross-section reaches the experiment bound when $|s_\theta|\to 1$ (compared to a similar issue concerning the $W$-mass anomaly and the EWPT $Z$-$Z'$ mixing discussed above), which requires a large $g,g_1$ coupling splitting such as $g_1=0.316g$, which is not signified as needing a study of the RGE, a task out of the scope of this work.    

\section{\label{con} Conclusion and outlook}

We have argued that there are two new physics phases associated with the 3-3-1 model, the 3-3-1 phase governed by a Higgs octet and the family-dependent abelian phase set by a Higgs singlet. This argument is valid for the 3-3-1 model with a high Landau pole like the 3-3-1 model with right-handed neutrinos considered in this work, while it is hard applying to the minimal 3-3-1 model, which has a Landau pole around 4--5 TeV.\footnote{See, for a recent study concerning Landau pole, \cite{barela}.}

Given that the Higgs octet VEV is much bigger than that of the Higgs singlet, the 3-3-1 phase is decoupled and imprinted at low energy being only a conserved matter parity. This matter parity may originate from a 3-3-1-1 symmetry as discussed in the paper or an alternative unification, which is left for a further study.\footnote{Flipped trinification \cite{ftn1,ftn2}, gauge family symmetry (see, e.g., \cite{fupov}), or a string model with different family constructions (see, for a review, \cite{string}) may be an alternative choice for unification.} This work simply supposes the matter parity working, neglecting its origin as well as any constraint from the more fundamental theory (if any).

Contrary to the usual 3-3-1 model studied at TeV scale as well as the scotogenic setup, the family-dependent abelian model instead yields compelling results for the $W$-mass anomaly, neutrino mass, and dark matter. Indeed, the $W$-mass measured signals a $Z$-$Z'$ mixing appropriate to the EWPT. The neutrino mass is induced by a scotogenic mechanism, naturally working due to the matter parity and isopartner presence. The model predicts a unique dark matter candidate, the light seesaw partner of the scotogenic Majorana field $N_R$. The family-dependent abelian phase is predicted at TeV scale and the large $SU(2)_L\otimes U(1)_{T_8}$ coupling-splitting is preferred.        

Who orders that: the third entries of lepton triplets? The present model reveals a possibility that the third entries of lepton triplets $N_{aL}$ contribute to WIMP dark matter, while their right-handed partners $N_{aR}$ contribute to scotogenic neutrino mass generation, which properly works if these $N$'s are electrically neutral and possess $B-L=0$. The phenomenology would differ from the current setup if either of the following cases occurs \ben \item $N_{aR}$ are suppressed: the scotogenic scheme is not viable, and needing a mechanism for $N_{aL}$ mass generation and that $N_{aL}$ are not necessarily the lightest dark isopartners; \item $n=2k/3\neq 0$ (still keeping $P=-1$ for $N$'s and other isopartners): we lack a mechanism for inducing Majorana $N_{aR}$ masses due to $B-L$ conservation, or otherwise the scotogenic setup does not work and $N$'s are dark Dirac fermions, not necessarily being the lightest dark isopartners. Additionally, $B-L$ gauge portal interaction for dark matter $N$ arises, although small; \item $n=(2k+1)/3$ (now $P=1$ for all fields): the lightest dark isopartner is not stabilized, i.e. the solution for dark matter disappears, besides the issue for Majorana $N_{aR}$ mass generation vs the scotogenic neutrino mass setup similar to the previous case, \item $N_{aL}=(\nu_{aR})^c$ whereas $N_{aR}$ are suppressed ($n=1$): in this case, $P=1$ is trivial for all fields and the scotogenic setup is not viable, i.e. both current schemes for neutrino mass and dark matter are discarded.\een In other words, all alternative choices of the third entires of lepton triplets lead to distinct phenomenologies for which the above four cases require different interpretations for neutrino mass and dark matter. Even, this is valid for the 3-3-1 model with $q=1$ in which the third entries contain $(e_{aR})^c$, or new charged leptons $E^+_a$ with $B-L=1$, or a new type of leptons with $B-L=n\neq 1$.


\begin{thebibliography}{99}

\bibitem{Kajita:2016cak} T. Kajita, Rev. Mod. Phys. {\bf 88}, 030501 (2016).

\bibitem{McDonald:2016ixn} A. B. McDonald, Rev. Mod. Phys. {\bf 88}, 030502 (2016).

\bibitem{Jungman:1995df} G. Jungman, M. Kamionkowski, and K. Griest, Phys. Rept. {\bf 267}, 195 (1996), arXiv:hep-ph/9506380 [hep-ph].

\bibitem{Bertone:2004pz} G. Bertone, D. Hooper, and J. Silk, Phys. Rept. {\bf 405}, 279 (2005), arXiv:hep-ph/0404175 [hep-ph].

\bibitem{Arcadi:2017kky} G. Arcadi, M. Dutra, P. Ghosh, M. Lindner, Y. Mambrini, M. Pierre, S. Profumo, and F. S. Queiroz, Eur. Phys. J. C {\bf 78}, 203 (2018), arXiv:1703.07364 [hep-ph].

\bibitem{CDF:2022hxs} T. Aaltonen {\it et al.} (CDF), Science {\bf 376}, 170 (2022).

\bibitem{331v1} M. Singer, J. W. F. Valle and J. Schechter, Phys. Rev. D {\bf 22}, 738 (1980). 

\bibitem{331v2} J. W. F. Valle and M. Singer, Phys. Rev. D {\bf 28}, 540 (1983). 

\bibitem{331pp} F. Pisano and V. Pleitez, Phys. Rev. {\bf D} 46, 410 (1992).

\bibitem{331f} P. H. Frampton, Phys. Rev. Lett. {\bf 69}, 2889 (1992).

\bibitem{331flt} R. Foot, H. N. Long, and Tuan A. Tran, Phys. Rev. D {\bf 50}, R34 (1994).  

\bibitem{ecq1} F. Pisano, Mod. Phys. Lett. A {\bf 11}, 2639 (1996). 

\bibitem{ecq2} A. Doff and F. Pisano, Mod. Phys. Lett. A {\bf 14}, 1133 (1999). 

\bibitem{ecq3} C. A. de S. Pires and O. P. Ravinez, Phys. Rev. D {\bf 58}, 035008 (1998). 

\bibitem{ecq4} C. A. de S. Pires, Phys. Rev. D {\bf 60}, 075013 (1999). 

\bibitem{ecq5} P. V. Dong and H. N. Long, Int. J. Mod. Phys. A {\bf 21}, 6677 (2006).  

\bibitem{neu1} M. B. Tully and G. C. Joshi, Phys. Rev. D {\bf 64}, 011301 (2001). 

\bibitem{neu2} A. G. Dias, C. A. de S. Pires, and P. S. R. da Silva, Phys. Lett. B {\bf 628}, 85 (2005). 

\bibitem{neu3} D. Chang and H. N. Long, Phys. Rev. D {\bf 73}, 053006 (2006). 

\bibitem{neu4} P. V. Dong, H. N. Long, and D. V. Soa, Phys. Rev. D {\bf 75}, 073006 (2007). 

\bibitem{neu5} P. V. Dong and H. N. Long, Phys. Rev. D {\bf 77}, 057302 (2008). 

\bibitem{neu6} P. V. Dong, L. T. Hue, H. N. Long, and D. V. Soa, Phys. Rev. D {\bf 81}, 053004 (2010). 

\bibitem{neu7} P. V. Dong, H. N. Long, D. V. Soa, and V. V. Vien, Eur. Phys. J. C {\bf 71}, 1544 (2011). 

\bibitem{neu8} P. V. Dong, H. N. Long, C. H. Nam, and V. V. Vien, Phys. Rev. D {\bf 85}, 053001 (2012). 

\bibitem{neu9} S. M. Boucenna, S. Morisi, and J. W. F. Valle, Phys. Rev. D {\bf 90}, 013005 (2014). 

\bibitem{neu10} S. M. Boucenna, R. M. Fonseca, F. Gonzalez-Canales, and J. W. F. Valle, Phys. Rev. D {\bf 91}, 031702 (2015). 

\bibitem{neu11} S. M. Boucenna, J. W. F. Valle, and A. Vicente, Phys. Rev. D {\bf 92}, 053001 (2015). 

\bibitem{neu12} H. Okada, N. Okada, and Y. Orikasa, Phys. Rev. D {\bf 93}, 073006 (2016). 

\bibitem{neu13} C. A. de S. Pires, Phys. Int. 6, 33 (2015).

\bibitem{d1} D. Fregolente and M.D. Tonasse, Phys. Lett. B {\bf 555}, 7 (2003). 

\bibitem{d2} H. N. Long and N. Q. Lan, Europhys. Lett. {\bf 64}, 571 (2003). 

\bibitem{d3} S. Filippi, W. A. Ponce, and L. A. Sanches, Europhys. Lett. {\bf 73}, 142 (2006). 

\bibitem{d4} C. A. de S. Pires and P. S. Rodrigues da Silva, J. Cosmol. Astropart. Phys. {\bf 12}, 012 (2007). 

\bibitem{d5} J. K. Mizukoshi, C. A. de S. Pires, F. S. Queiroz, and P. S. Rodrigues da Silva, Phys. Rev. D {\bf 83}, 065024 (2011). 

\bibitem{d6} J. D. Ruiz-Alvarez, C. A. de S. Pires, F. S. Queiroz, D. Restrepo, and P. S. Rodrigues da Silva, Phys. Rev. D {\bf 86}, 075011 (2012). 

\bibitem{d7} P. V. Dong, T. P. Nguyen, and D. V. Soa, Phys. Rev. D {\bf 88}, 095014 (2013). 

\bibitem{d8} S. Profumo and F.S. Queiroz, Eur. Phys. J. C {\bf 74}, 2960 (2014). 

\bibitem{d9} C. Kelso, C. A. de S. Pires, S. Profumo, F. S. Queiroz, and P. S. Rodrigues da Silva, Eur. Phys. J. C {\bf 74}, 2797 (2014). 

\bibitem{d10} P. S. Rodrigues da Silva, Phys. Int. {\bf 7}, 15 (2016).

\bibitem{d11} P. V. Dong, N. T. K. Ngan, and D. V. Soa, Phys. Rev. D {\bf 90}, 075019 (2014). 

\bibitem{d12} P. V. Dong, C. S. Kim, N. T. Thuy, and D. V. Soa, Phys. Rev. D {\bf 91}, 115019 (2015).

\bibitem{d13} P. V. Dong, T. D. Tham, and H. T. Hung, Phys. Rev. D {\bf 87}, 115003 (2013). 

\bibitem{d14} P. V. Dong, D. T. Huong, F. S. Queiroz, and N. T. Thuy, Phys. Rev. D {\bf 90}, 075021 (2014). 

\bibitem{d15} D. T. Huong, P. V. Dong, C. S. Kim, and N. T. Thuy, Phys. Rev. D {\bf 91}, 055023 (2015). 

\bibitem{d16} D. T. Huong and P. V. Dong, Eur. Phys. J. C {\bf 77}, 204 (2017).

\bibitem{d17} A. Alves, G. Arcadi, P. V. Dong, L. Duarte, F. S. Queiroz, and J. W. F. Valle, Phys. Lett. B {\bf 772}, 825 (2017). 

\bibitem{d18} P. V. Dong, D. T. Huong, D. A. Camargo, F. S. Queiroz, and J. W. F. Valle, Phys. Rev. D {\bf 99}, 055040 (2019).

\bibitem{wm331} D. V. Loi and P. V. Dong, Eur. Phys. J. C {\bf 83}, 56 (2023).

\bibitem{ano1} D. J. Gross and R. Jackiw, Phys. Rev. D {\bf 6}, 477 (1972). 

\bibitem{ano2} H. Georgi and S. L. Glashow, Phys. Rev. D {\bf 6}, 429 (1972). 

\bibitem{ano3} J. Banks and H. Georgi, Phys. Rev. D {\bf 14}, 1159 (1976). 

\bibitem{ano4} S. Okubo, Phys. Rev. D {\bf 16}, 3528 (1977).

\bibitem{queiroz} A. Alves, L. Duarte, S. Kovalenko, Y. M. Oviedo-Torres, F. S. Queiroz, and Y. S. Villamiza, Phys. Rev. D {\bf 106}, 055027 (2022).

\bibitem{mazeronu} E. Ma, Phys. Rev. Lett. {\bf 86}, 2502 (2001).

\bibitem{mp2} P. V. Dong, Phys. Rev. D {\bf 92}, 055026 (2015).

\bibitem{dongfc} P. V. Dong, T. N. Hung, and D. V. Loi, Eur. Phys. J. C {\bf 83}, 199 (2023).

\bibitem{km2022} M. Bauer and P. Foldenauer, Phys. Rev. Lett. {\bf 129}, 171801 (2022). 

\bibitem{stu} M. E. Peskin and T. Takeuchi, Phys. Rev. D {\bf 46}, 381 (1992).

\bibitem{strumia} A. Strumia, JHEP {\bf 08}, 248 (2022).

\bibitem{zzprimemixing} J. Erler, P. Langacker, S. Munir, and E. Rojas, JHEP {\bf 08}, 017 (2009).

\bibitem{fcnceff} M. Bona {\it et al.} (UTfit), JHEP {\bf 03}, 049 (2008). 

\bibitem{fcnceff1} G. Isidori, Y. Nir, and G. Perez, Ann. Rev. Nucl. Part. Sci. {\bf 60}, 355 (2010).

\bibitem{pdg} R.L. Workman {\it et al.} (Particle Data Group), Prog. Theor. Exp. Phys. {\bf 2022}, 083C01 (2022).

\bibitem{sgsetup} E. Ma, Phys. Rev. D {\bf 73}, 077301 (2006).

\bibitem{vdoublet} P. V. Dong, D. V. Loi, L. D. Thien, and P. N. Thu, Phys. Rev. D {\bf 104}, 035001 (2021).

\bibitem{3311scoto} A. E. C. Hernandez, C. Hati, S. Kovalenko, J. W. F. Valle, and C. A. Vaquera-Araujo, JHEP {\bf 03}, 034 (2022). 

\bibitem{3311scoto1} A. E. C. Hernandez, J. W. F. Valle, and C. A. Vaquera-Araujo, Phys. Lett. B {\bf 809}, 135757 (2020).

\bibitem{3311scoto2} J. Leite, O. Popov, R. Srivastava, and J. W. F. Valle, Phys. Lett. B {\bf 802}, 135254 (2020). 

\bibitem{sddm} G. Belanger, F. Boudjema, A. Pukhov, and A. Semenov,
Comput. Phys. Commun. {\bf 180}, 747 (2009).

\bibitem{ddsd} LZ collaboration, arXiv:2207.03764 [hep-ex]; XENON collaboration, Phys. Rev. Lett. {\bf 122}, 141301 (2019); PandaX-II collaboration, Phys. Lett. B {\bf 792}, 193 (2019).

\bibitem{barela} M. W. Barela, arXiv:2305.05066 [hep-ph].

\bibitem{ftn1} D. T. Huong and P. V. Dong, Phys. Rev. D {\bf 93}, 095019 (2016).

\bibitem{ftn2} P. V. Dong, D. T. Huong, F. S. Queiroz, J. W. F. Valle, and C. A. Vaquera-Araujo, JHEP {\bf 04}, 143 (2018).

\bibitem{fupov} Z. G. Berezhiani and M. Y. Khlopov, Sov. J. Nucl. Phys. {\bf 51} 739 (1990).

\bibitem{string} P. Langacker and M. Plumacher, Phys. Rev. D {\bf 62} 013006 (2000).

\end{thebibliography}
\end{document}